\documentclass[]{spie}  
\usepackage[]{graphicx}
\usepackage{setspace}
\usepackage{tocloft}
\usepackage{verbatim}

\title{Characterization of the Atmospheric Dispersion Corrector of the Gemini Planet Imager} 

\author{Pascale Hibon\supit{a}, Sandrine Thomas\supit{b}, Jennifer
  Dunn\supit{c},  Jenny Atwood\supit{c}, Les Saddlemyer\supit{c}, Naru
  Sadakuni\supit{a}, Stephen Goodsell\supit{d}, Bruce Macintosh\supit{e}, James Graham\supit{h}, Marshall Perrin\supit{f},
  Fredrik Rantakyro\supit{a}, Vincent Fesquet \supit{a},
 Andrew Serio\supit{a}, Carlos Quiroz\supit{a}, Andrew Cardwell\supit{a},
  Gaston Gausachs\supit{a}, Dmitry Savransky\supit{g}, Dan Kerley\supit{c}, Markus Hartung\supit{a}, Ramon Galvez\supit{a} and Kayla Hardie\supit{a}.
\skiplinehalf
\supit{a}Gemini South Observatory, Casilla 603, La Serena, Chile\\
\supit{b}NASA Ames Research Center, MS 244-10, Moffett Field, CA 94035, USA\\
\supit{c} NRC Herzberg Institute of Astrophysics, 5071 West Saanich Road, Victoria, British Columbia, V9E 2E7 Canada \\
\supit{d} Gemini North Observatory,670 N. A’ohoku Place,Hilo, HI\\
\supit{e} Physics and Astrophysics Bldg., Rm. 215, Physics Dept., Stanford University, Stanford, CA 94305\\
\supit{f} The Space Telescope Science Institute, 3700 San Martin Drive, Baltimore, MD 21218, MD, U.S.\\
\supit{g} Sibley School of Mechanical and Aerospace Engineering, Cornell University, Ithaca, NY 96720 USA\\
\supit{h} Department of Astronomy, UC Berkeley, Berkeley CA 94720, USA
}

\authorinfo{ Address all correspondence to : Pascale Hibon, Gemini South Observatory, Casilla 603, La Serena, Chile, Tel: +56 512-205-680; E-mail: phibon@gemini.edu} 

\begin{document} 
\maketitle 

\begin{abstract}
An Atmospheric Dispersion Corrector (ADC) uses a double-prism arrangement to nullify the vertical chromatic dispersion introduced by the atmosphere at non-zero zenith distances. \\
The ADC installed in the Gemini Planet Imager (GPI) was first tested in August 2012 while the instrument was in the laboratory. GPI was installed at the Gemini South telescope in August 2013 and first light occurred later that year on November 11th. \\
In this paper, we give an overview of the characterizations and performance of this ADC unit obtained in the laboratory and on sky, as well as the structure of its control software.

\end{abstract}




\section{Introduction}
\label{sect:intro}  

\subsection{Definition and Role of the Atmospheric Dispersion Corrector}
\begin{figure}[!h]
   \begin{center}
   \begin{tabular}{c c}
   \includegraphics[height=5cm]  {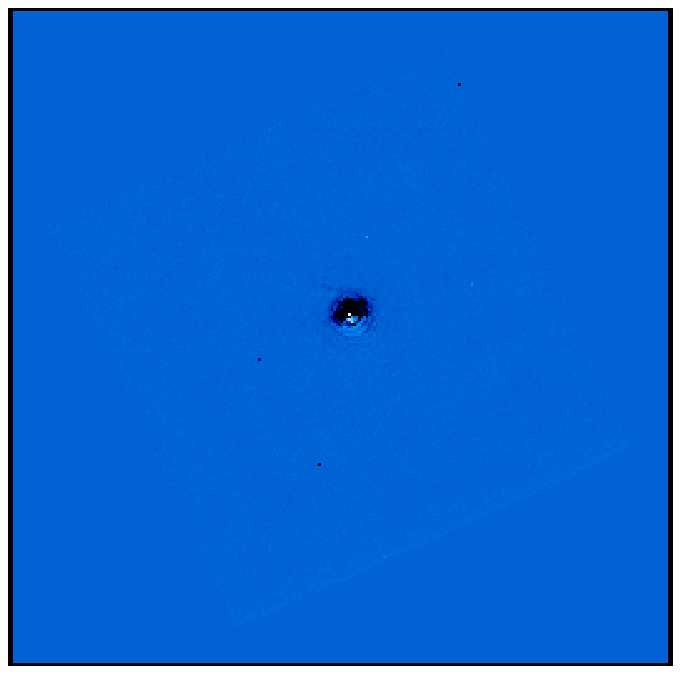} & \includegraphics[height=5cm]  {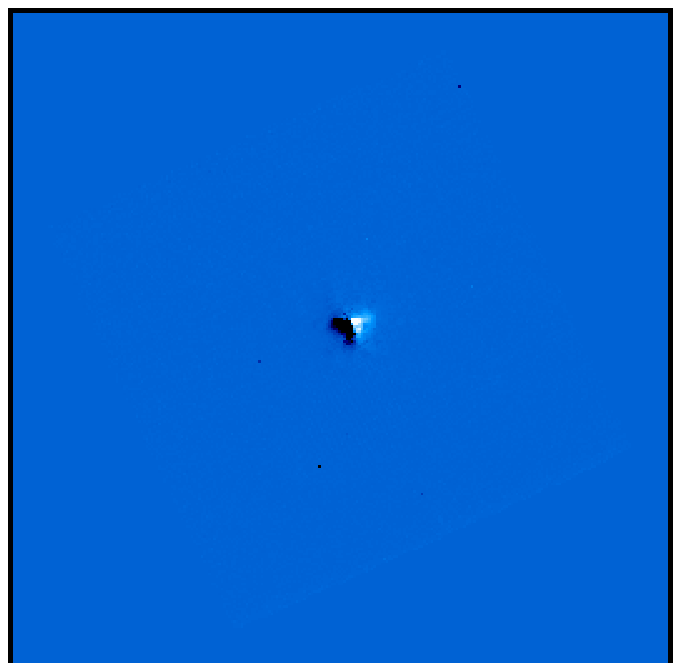}\\
\textit{a. ADC deployed} & \textit{b. ADC extracted}
   \end{tabular}
   \end{center}
   \caption 
   { \label{dispex} 
Visualization of the effect of the presence/absence of the ADC. These images show the difference between the first and last wavelength slice of datacubes taken in J-band. Figure a. shows this difference in the case of a deployed ADC. Figure b. in the case of an extracted ADC.} 
   \end{figure}

Light that enters our atmosphere at an angle relative to the local vertical will suffer refraction which bends the light.
The refractive (bending) power of the atmosphere is wavelength dependent (called dispersion) which means that the deviation that atmospheric refraction causes depends not only on the angle of the light but also the colour of that light (see Figure~\ref{dispex}). This phenomenon of atmospheric dispersion spreads the light from a point source into a spectrum of colours and this spread is greater at lower elevation. To search for exoplanets and disks, a broad band performance is not necessary given the coronagraph chromaticity.

\subsection{GPI ADC}
The Gemini Planet Imager is a high contrast instrument dedicated
to the direct imaging and the integral field spectroscopic
characterizing of extrasolar planets and planetary debris disks\cite{macintosh2014first}. It was installed at the Gemini
South Telescope in August 2013 and saw its first light in November
2013 \cite{rantakyrothis}.\\
 Differential Atmospheric Refraction
effects are a lot less important inside a small
bandpass. Since more than 90\% of GPI observation time will be dedicated
to the search of exoplanets and disks, having a removable/deployable ADC
is logical. \\
- The chromatic dispersion should be minimized at the focal plane such
that no star light at any detectable wavelength reflects into the
Integral Field Spectrograph (IFS).\\
- Minimizing the dispersion reduces non-common path errors. This
increases the AO systems ability to correct.  

\subsubsection{ADC optical design}
\begin{figure}[!h]
   \begin{center}
   \begin{tabular}{c}
   \includegraphics[height=7cm]  {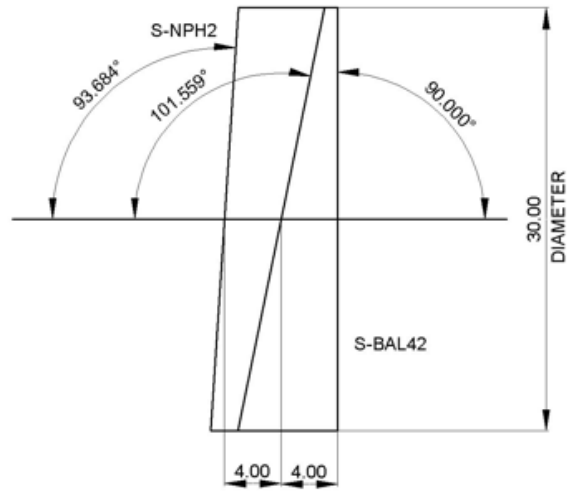}
   \end{tabular}
   \end{center}
   \caption 
   { \label{design} 
A completed doublet,  one of two which comprise the GPI ADC. Linear units in millimeters. } 
   \end{figure}

The ADC optical design selected for GPI is a doublet linear ADC. Another common design is a so-called Risley prism. A comparison of the performance of a doublet linear ADC and a Risley ADC in the case of GPI are described in Table~\ref{diff}. 
The doublet linear ADC design offers fewer degrees of motion and the absence of non-common path errors.
A doublet linear ADC uses a two identical prism doublets where the effective prism angle
can be varied by altering the relative orientation between the two
prisms.  
One element of the doublet is made of S-NPH2, the other of S-BAL42 (Ohara glass catalog),  as seen in
Figure~\ref{design}. Each doublet rotates on the optical axis to
adjust the wedge with
  the dispersion direction. One prism doublet translates along the
  optical axis to vary the dispersion
  correction. 
In doing this one can set the device so that the dispersion it introduces nullifies the vertical dispersion introduced by the atmosphere. These different movements
provide an adjustable correction for any altitude. 
It is important to realise that the ADC needs to actually introduce a
reconvergence of the angularly separated colour images so they all
meet up again at the focal plane.
The GPI ADC is bolted to the mounting structure (see Figure~\ref{mecha}) and is located just behind the entrance window.
The properties of the ADC derived from GPI's science requirements are listed in Table~\ref{require}.

\subsubsection{ADC Mechanical design and alignment.}
\begin{figure}[!h]
   \begin{center}
   \begin{tabular}{c}
   \includegraphics[height=5.5cm]  {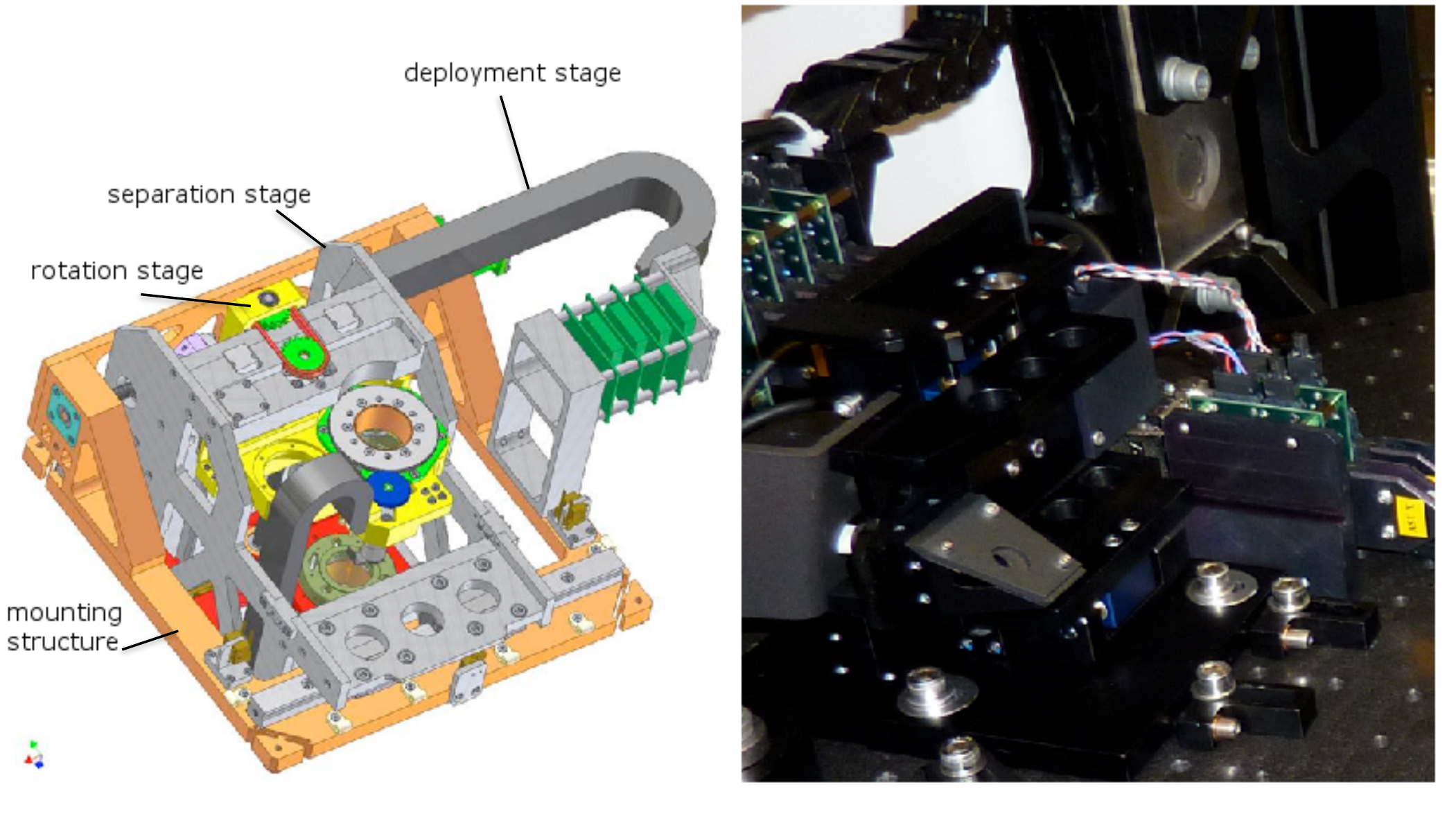}
   \end{tabular}
   \end{center}
   \caption 
   { \label{mecha} 
Atmospheric Dispersion Unit Mechanical drawing and mechanism.} 
   \end{figure}

The mechanical design of the ADC must control various degrees of freedom of the  two prism doublets that comprise the ADC assembly. These degrees of freedom include: prism absolute rotation, prism relative rotation, individual prism element tip/tilt, prism Group Tip/Tilt, and prism separation. 
Considering both prism doublets as a single rigid body, the axis connecting their geometric centers must be aligned to the optical axis of the GPI instrument. To eliminate optical ghosts, the prism doublets are tilted in a direction which is mutually orthogonal to the optical axis and to each other. Both prisms are subject to the primary tilt; only one prism receives the secondary tip.
To provide the appropriate degree of atmospheric compensation three active adjustments are necessary. First, each prism doublet must be rotated to align the direction of the wedge angle with the direction of optical dispersion induced by the atmosphere. Second, one prism doublet must be rotated to align the direction of the wedge angle relative to the second prism doublet. This relative alignment specification applies over the entire range of motion. Third, to control the amount of dispersion correction, the separation distance between the inside faces of the prism doublets must be varied.
Consequently, the ADC has been designed as a four-axis mechanism, with one linear stage for deployment into the optical path, one linear stage for controlling the separation distance between prisms, and two identical but independent rotational stages for rotating the prisms about the optical axis (see Figure~\ref{mecha}).  

\subsubsection{ADC software system}
GPI's command structure is
hierarchical and the GPI Top Level Computer (TLC) \cite{dunnthis,rantakyrothis} 
 is the one that
initiates commands.  All commands coming
from Gemini are sent via the Gemini Instrument Application Programmer
Interface (GIAPI).  The commands all go through the Instrument
Sequencer, which is the software process that coordinates activities
within GPI.  The ADC Assembly is a layer of abstraction that allows
the commanding of the ADC in user units or alternatively by providing
a Cassegrain  or zenith distance.  The final layer is the Motion Control
Daemon which controls the individual motors by commanding them in mm
and degrees.  GPI has a flexible architecture that allows the
control at various different software layers. Setup of the software system includes the
determination if atmospheric correction is required.  
The three subsystems of GPI, AOC (Adaptiv Optics Computer), CAL (Calibration) and Integral Field Spectrograph (IFS) , each operate over different wavelength bands.  Differential atmospheric refraction (DAR) effects will therefore result in a given image appearing in slightly different locations in the three subsystems.
GPI can be operated either with the ADC deployed or retracted. If the ADC is retracted then the DAR effects will be larger than the deployed case. Adjustments to how the light is directed through GPI is done using pointing and centering mirrors. \\
The ADC orientation angle and separation distance can be set directly to provide flexibility during development.  During operation the ADC's orientation and/or separation is based on the current telescope Cassegrain and zenith distances. 
Calculations to determine the ADC separation (dispersion correction
power) involves:\\
- Using the current temperature, pressure and humidity at the telescope to determine the refractive index of the air and dispersion in the science band.\\
- The absolute refraction and the true zenith distance are calculated using the zenith distance and index of refraction\\
- A Zemax model was used to determine initial dispersion
conversions but the coefficients can be replaced once on sky
calibration is completed.\\
The dispersion in the calibration and AO bands are also calculated in order to determine expected image shifts between GPI subsystems. The ADC orientation is determined from the telescope Cassegrain and parallactic angles.  Each prism is offset from the axis of dispersion by equal and opposite amounts.\\

The purpose of this study is to characterize the ADC installed in GPI. We will present in Section~\ref{labo} the characterization in the laboratory in the University of California, Santa Cruz and in Section~\ref{onsky}, the results obtained on-sky in terms of contrast, throughput and astrometry.

  \begin{figure}[!h]
   \begin{center}
   \begin{tabular}{c}
   \includegraphics[height=7cm]  {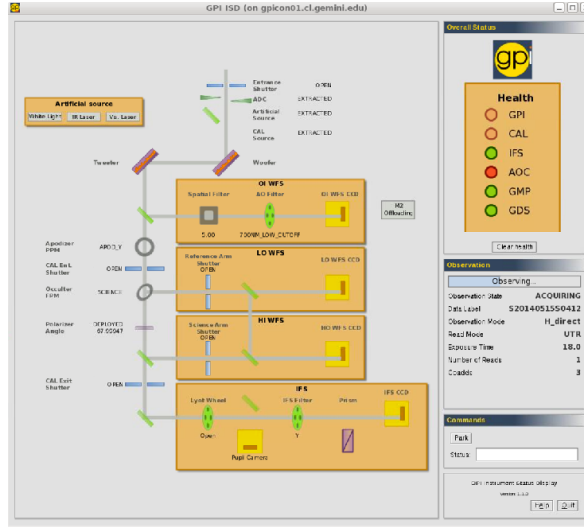}
   \end{tabular}
   \end{center}
   \caption 
   { \label{optlayout} 
Optical layout of GPI, showing the Artificial Source being downstream
of the ADC. The light path inside the instrument is also marked.} 
  \end{figure}


\begin{table}[h]
\caption{Adopted ADC requirements.} 
\label{require}
\begin{center}       
\begin{tabular}{|l|l|} 
\hline
\rule[-1ex]{0pt}{3.5ex}  Temperature & -5C to 20C  \\
\hline\hline
\rule[-1ex]{0pt}{3.5ex} Humidity & 0- 100\%  \\
\hline
\rule[-1ex]{0pt}{3.5ex}  Altitude &0-15.5km   \\
\hline
\hline
\rule[-1ex]{0pt}{3.5ex}  Wavelength range & $0.7\leq \lambda \leq
2.19\mu m$ in passbands \textit{Y, J, H ,K1, K2}.  \\
\hline
\rule[-1ex]{0pt}{3.5ex}  Residual dispersion & $\leq$5mas at the focal
plane.\\
\hline
\rule[-1ex]{0pt}{3.5ex}  Zenith Distance & $1 \leq ZD \leq 50$deg  \\
\hline
\rule[-1ex]{0pt}{3.5ex}  Field of view & $2.78 \times 2.78$ arcsec \\
\hline
\end{tabular}
\end{center}
\end{table} 


\begin{table}[h]
\caption{Performance comparison between a doublet linear ADC and a
  Risley ADC.} 
\label{diff}
\begin{center}       
\begin{tabular}{|l|l|l|} 
\hline
\rule[-1ex]{0pt}{3.5ex} & Doublet Linear  & Risley  \\
\hline\hline
\rule[-1ex]{0pt}{3.5ex} Observing bands & \textit{Y, J, H, K} & \textit{Y, J, H, K}\\
\hline
\rule[-1ex]{0pt}{3.5ex}  Prism material & S-NPH2/S-BAL42 & BaF2/CaF2/S-NPH2 \\
\hline
\rule[-1ex]{0pt}{3.5ex} Optical transmission at 1$\mu$m & 99.44\%& 95.6\% \\
\hline
\rule[-1ex]{0pt}{3.5ex}  Optical transmission at 2.4$\mu$m & 85.85\%& 92.3\% \\
\hline
\end{tabular}
\end{center}
\end{table} 

\section{Laboratory Characterization} \label{labo}
\subsection{Initial configuration}	

Since the Artificial Source Unit (ASU) is downstream of the ADC (see Figure~\ref{optlayout}), we used a telescope simulator external to GPI for the experiment. The tests were done in H band.\\
Because the telescope simulator only provides a point source instead of an object dispersed by the atmosphere as an  input, tests results in the laboratory will not reproduce the same configuration as on the telescope. The method here is to measure how much dispersion the ADC introduces and compare it with theoretical models. \\
Another difference between laboratory and sky procedures is the correction of the tilt introduced by the ADC and the correction of the centering. In the laboratory, we use the woofer to correct for the tilt. On sky, the tilt is corrected by re-pointing the telescope and the centering is corrected by the input fold. Any errors or aberrations induced by the telescope or the ADC due to the object being off-axis, will not be captured in these tests.
When deployed, the ADC should not introduce any vignetting and should deliver the best image quality with a minimum target displacement. 
The ADC prisms are on independent rotation stages, allowing them to be rotated relative to each other or as a block. Both degrees of freedom allow to find the deploy position that will match the requirement as well as the Zemax model. 
Moreover,  The whole ADC assembly rotates and will need to be aligned with the parallactic angle, which determines the
direction of atmospheric dispersion. After some tests, the software was updated with offsets at the assembly
level to move to the deploy position, aligned with the OMSS axis.

\subsection{Search for the initial deployed position}
\begin{figure}[!h]
 \centering
  \includegraphics[width=10cm]{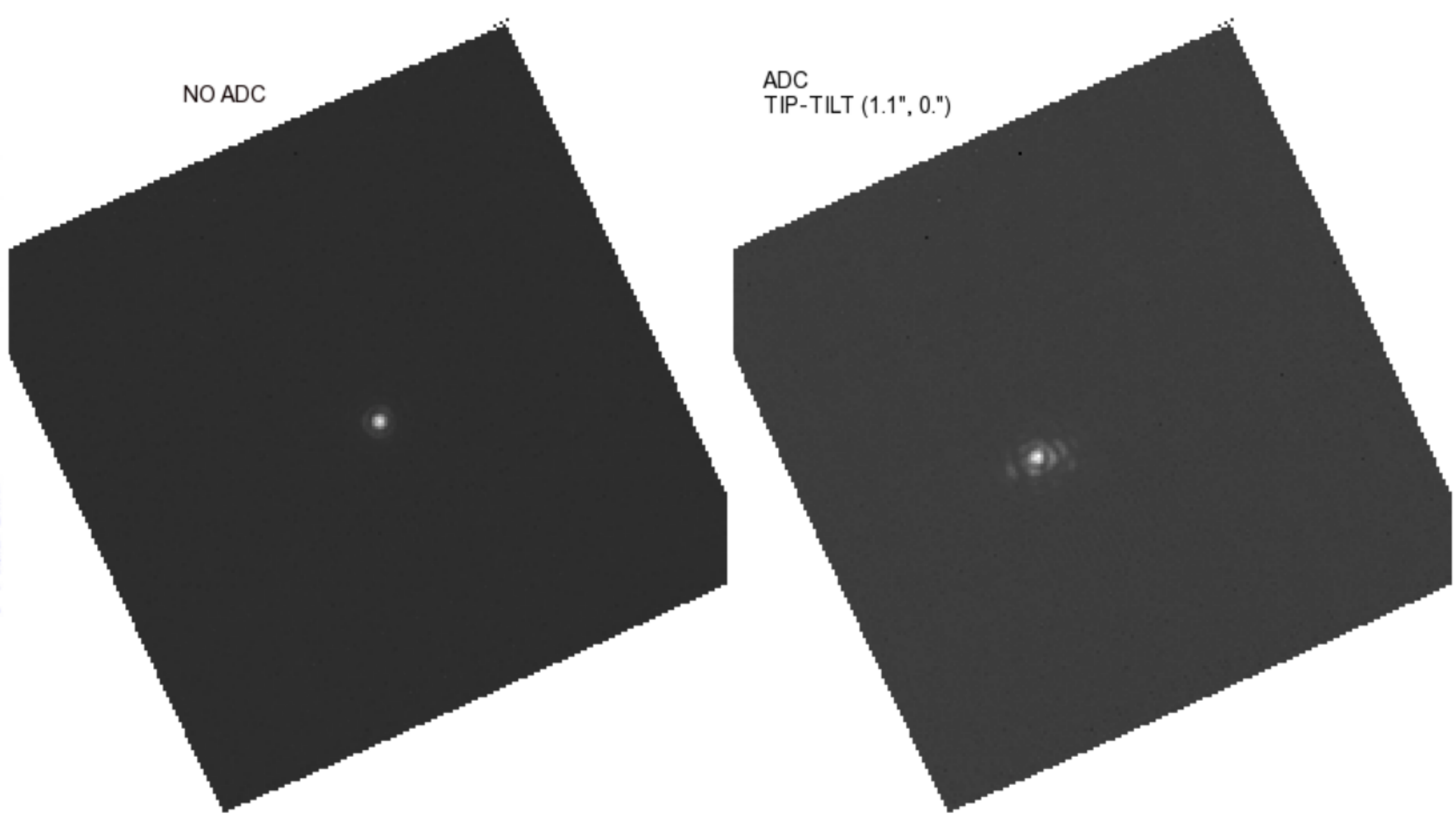}
  \caption{Image taken in laboratory with and without the ADC deployed. We can observe that no aberration have been introduced by the ADC once deployed. Also, one can see the offset of the PSF. }
  \label{deploy}
\end{figure}

When deployed, the ADC should not introduce any vignetting and should deliver the best image quality with a minimum target displacement. \\
The ADC prisms are on independent rotation stages, allowing them to be rotated relative to each other or as a block. Both degrees of freedom allow to find the deploy position that will match the requirement as well as the Zemax model. 
Moreover,  the whole ADC assembly rotates and will need to be aligned with the parallactic angle, which determines thedirection of atmospheric dispersion. After some tests, the software was updated with offsets at the assembly
level to move to the deploy position, aligned with the OMSS axis.\\
Due to errors in the manufacture of the ADC, the
beam exits the prisms at an angle. 
This causes pointing and centering errors (see Figure~\ref{deploy}). GPI is equipped with an input fold located after the input focal plane of the instrument. Its function is to align the pupil of the telescope relative to the MEMS and thus to all pupils inside GPI. Since the input fold is not in a focal plane, changing its position will add an offset to the pointing introduced by the ADC. The telescope will correct for that final pointing error.

\subsection{Translation - Dispersion}
\begin{figure}[!h]
 \centering
  \includegraphics[width=11cm]{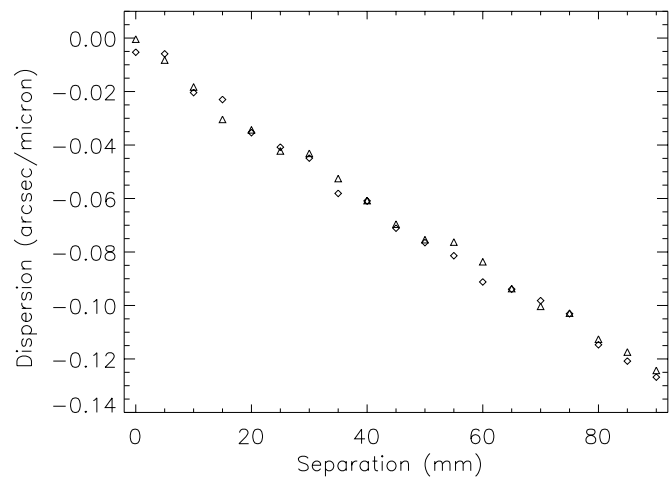}
  \caption{Repeatibility of the ADC dispersion performance while increasing or
    decreasing the separation between the two prism doublets. The
    triangles represent the dispersion values obtained when increasing
  the separation and the diamonds while decreasing it.}
  \label{repeat}
\end{figure}

This test consists of determining the dispersion we observe in
the IFS images while increasing/decreasing the distance between the
two prisms. The distance between two prisms varies from 0mm to 90mm.  \\
The general behavior of the ADC while increasing and decreasing the
separation by 5mm increments from 0mm to 90mm (and from 90mm to 0mm) between the two prism doublets is shown in
Figure~\ref{repeat}. These data were taken using the
H-band direct observing mode with the artificial source. The exposure
time was set to 1.5s. The peak-to-valley dispersion was then measured in each cube and is represented by diamonds (in the case of decreasing the separation) and triangles (in the case of increasing the separation) in Figure~\ref{repeat}.\\
These data also allow us to test the repeatability of the ADC dispersion
performance.
A seen in the Figure~\ref{repeat}, the dispersion value, at a same
separation, is very similar. This shows then that the ADC system is reliable.

\begin{figure}[!h]
 \centering
  \includegraphics[width=11cm]{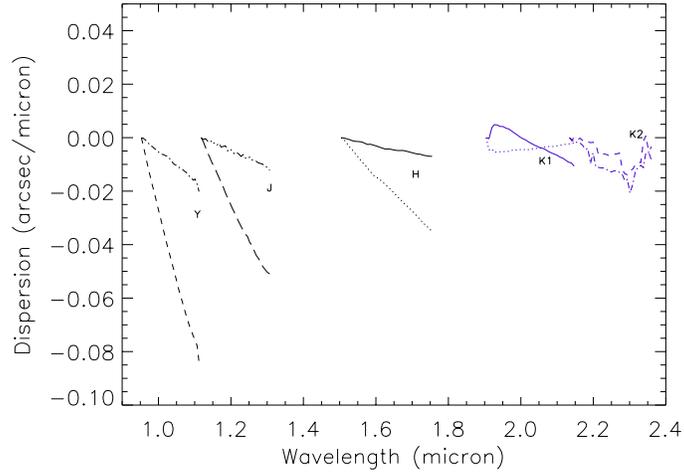}
  \caption{Dispersion in each filter in the case of the minimum and
    maximum separation between the two prism doublets. Each set of two
  lines represents one filter. For the Y-band filter : The dot-dashed
  line is the minimum separation, 5mm, and the dashed line is the maximum
  separation, 90mm. For J-band filter : the 3 dot-dashed is the minimum
  separation, the long-dashed the maximum separation. For the H-band
  filter : the solid line is the minimum, the dotted line is the
  maximum. For the K1-band filter : the blue dotted line is the
  minimum and the blue solid line is the maximum. For the K2-band
  filter : the blue long-dashed line is the minimum and the blue
  dot-dashed line the maximum. This allows us
  to visualize the dispersion range available in each filter. The wavelength solutions for K1- and K2-bands were not optimal, which explains the behavior of the curves in this filter.} 
  \label{allfilt}
\end{figure}

In order to test the atmospheric dispersion in function of the
wavelength, we took data at 5mm and 90mm separation, two extreme
separation values for the prism doublets, for Y, J, H, K1
and K2-band filters. Figure~\ref{allfilt} shows the dispersion
obtained in these data. 

\begin{figure}[hp]
\begin{center}
\begin{tabular}{c c}
\includegraphics[scale=0.25]{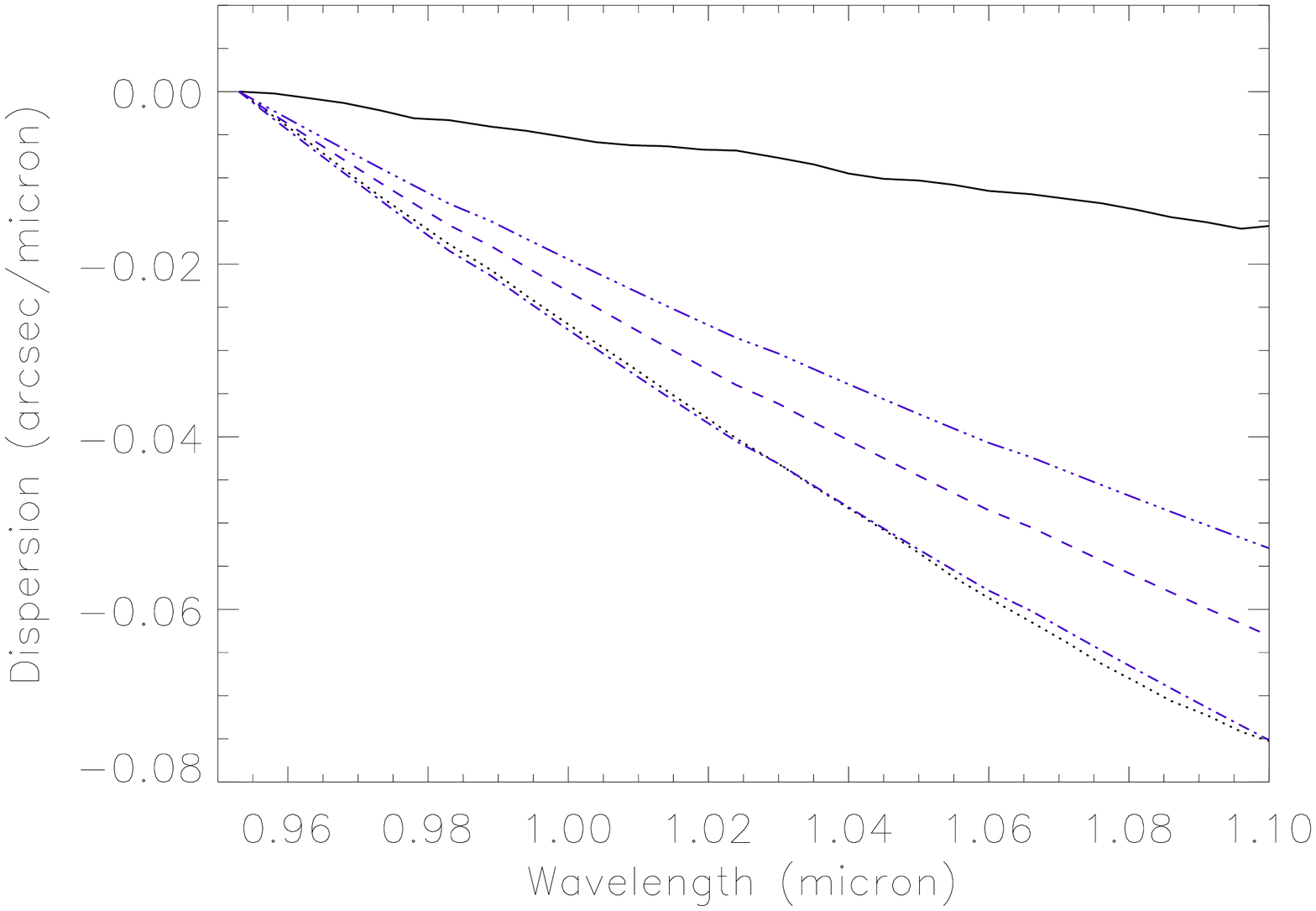}
&\includegraphics[scale=0.25]{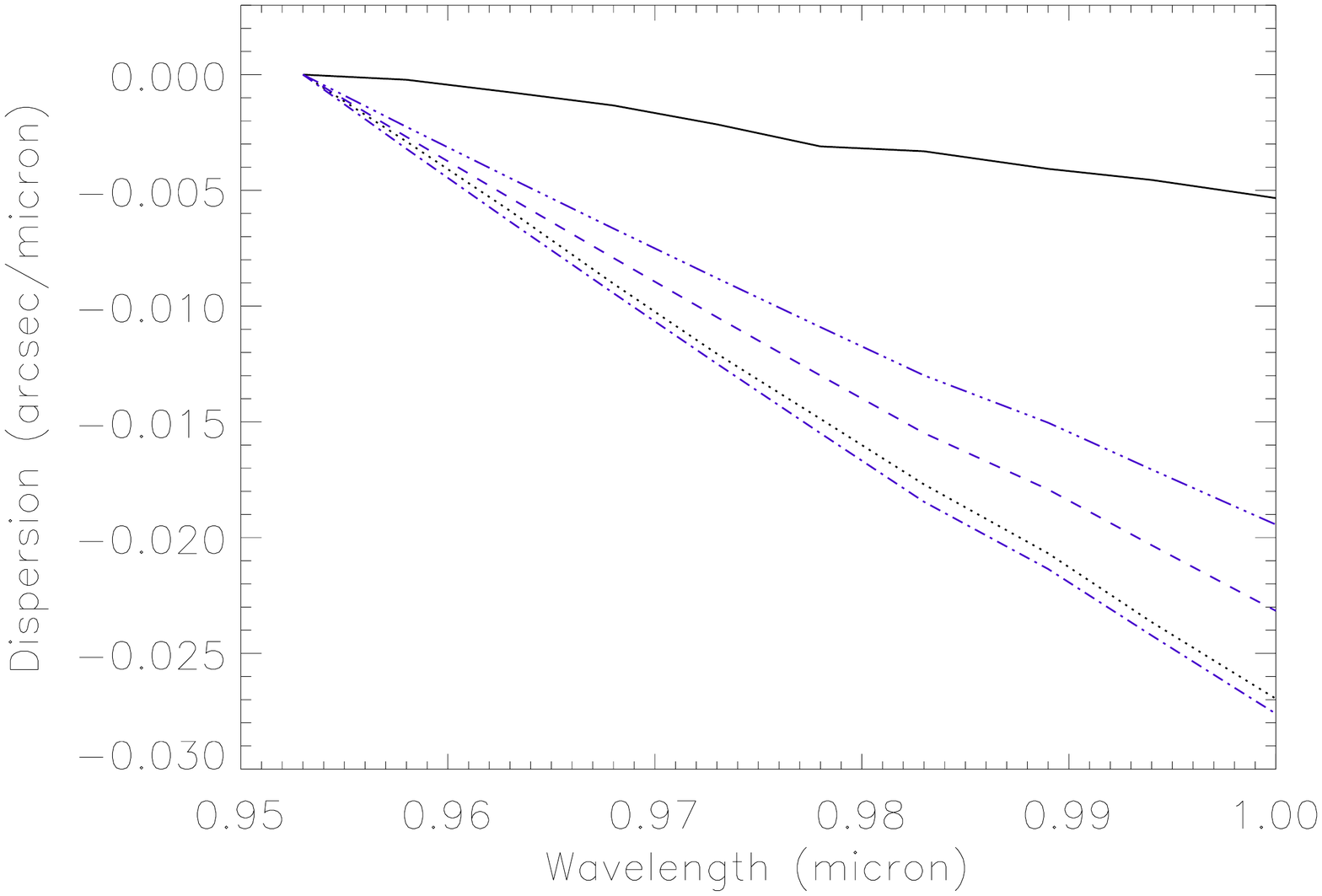}\\
\textit{a.Y-band} & \textit{b. Y-band zoom}\\
\includegraphics[scale=0.25]{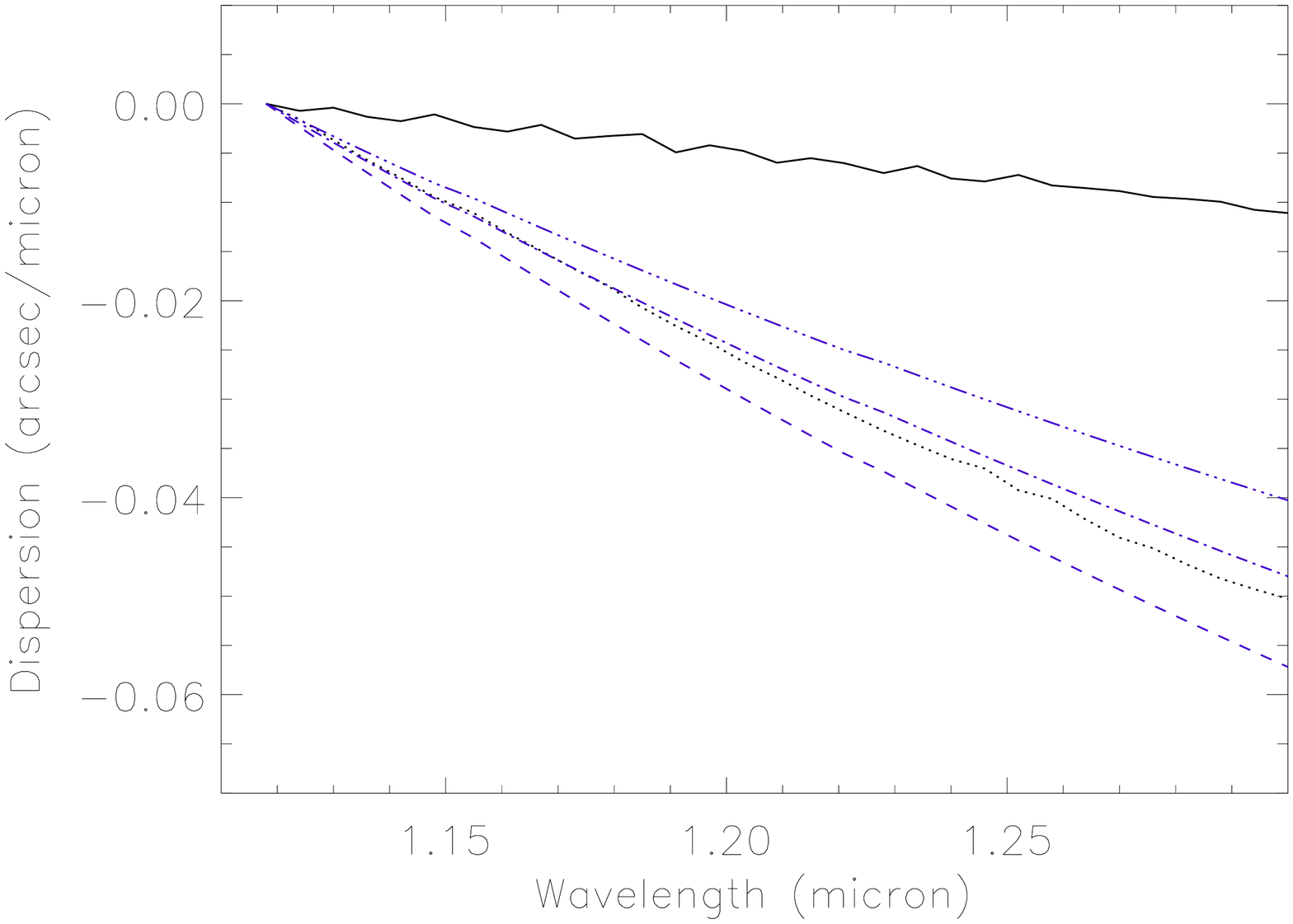}
&\includegraphics[scale=0.25]{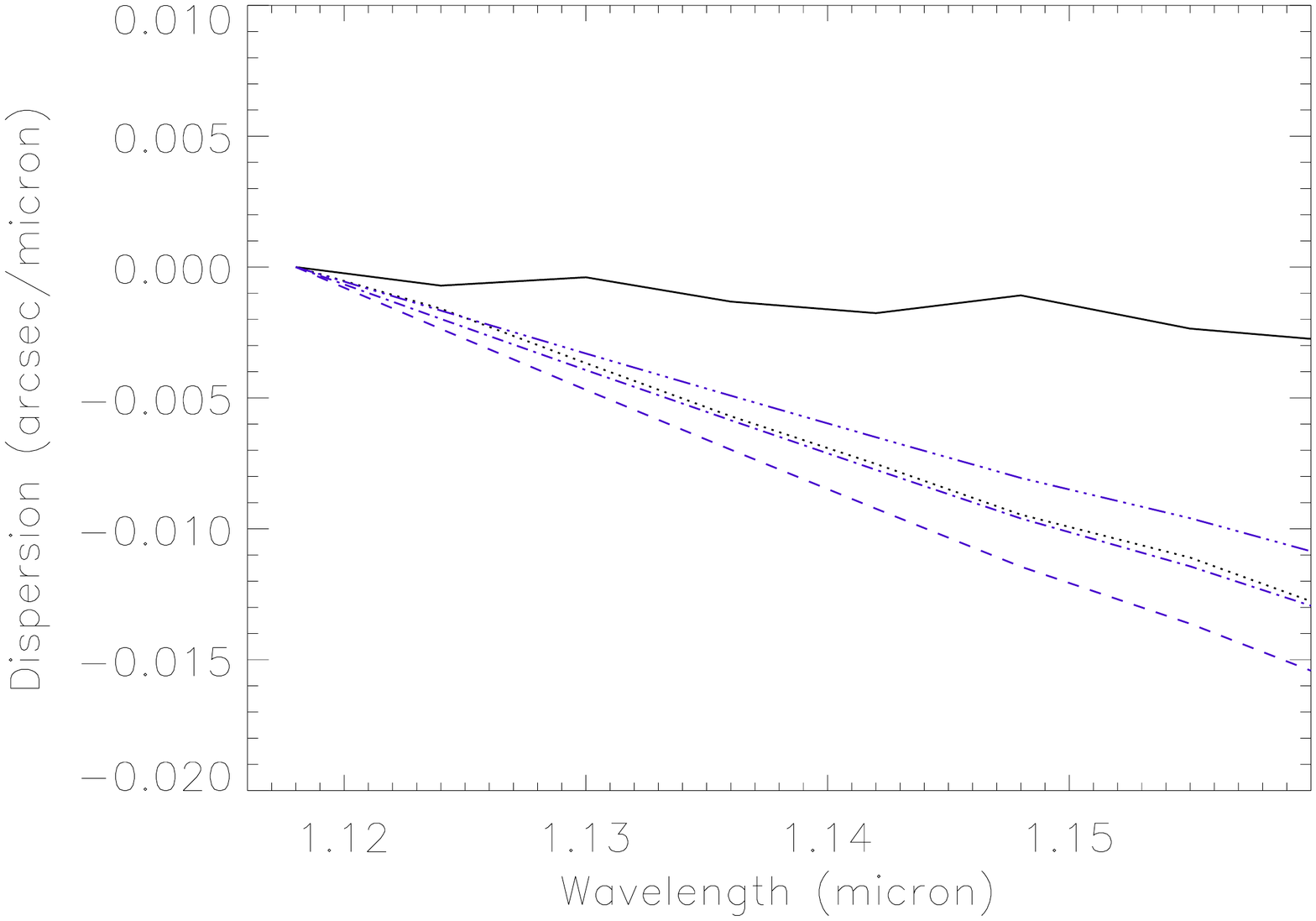}\\
\vspace{-0.5mm}
\textit{c.J-band} & \textit{d. J-band zoom}\\
\includegraphics[scale=0.25]{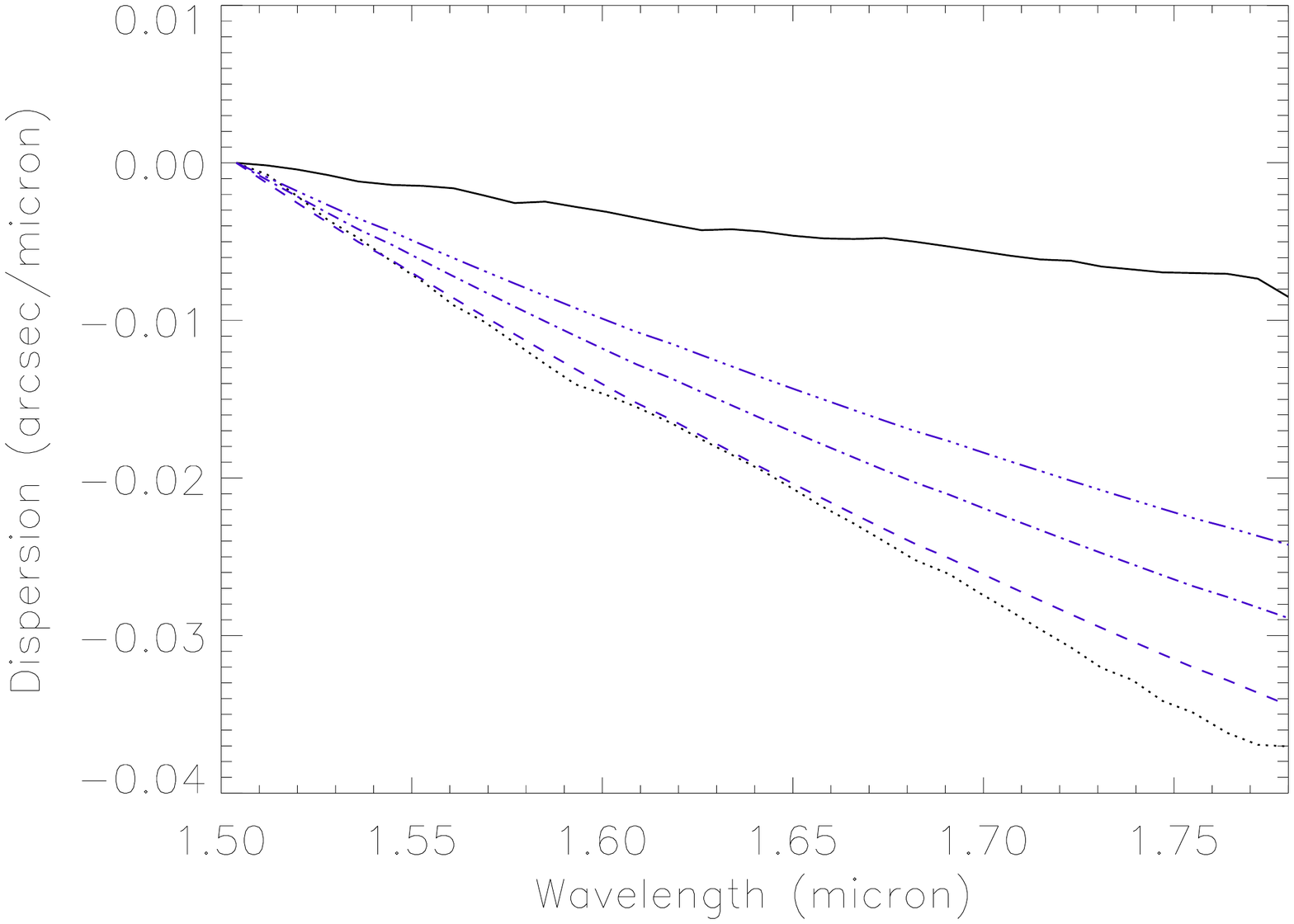} &\includegraphics[scale=0.25]{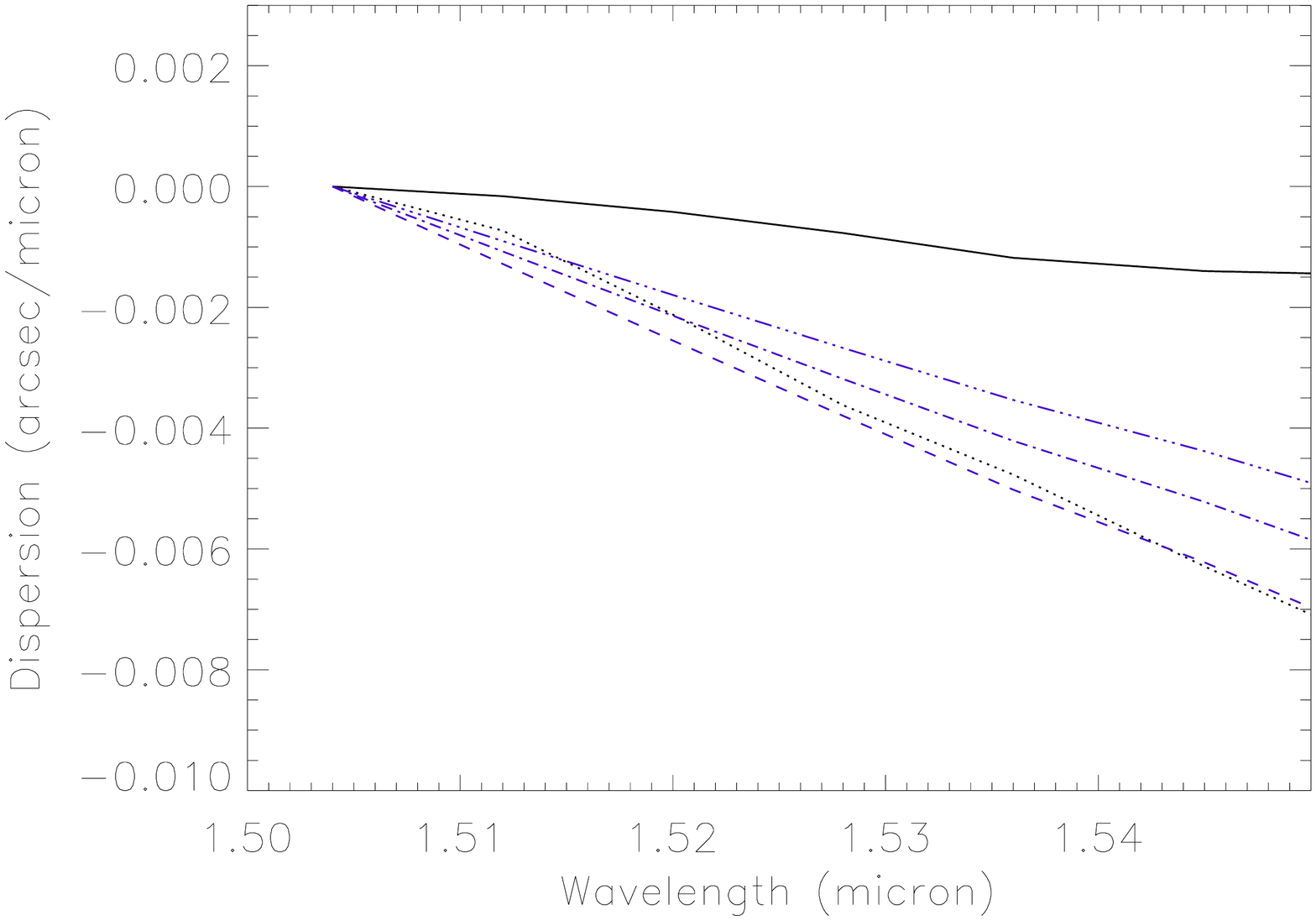}\\
\textit{e.H-band} & \textit{f. H-band zoom}\\
\includegraphics[scale=0.25]{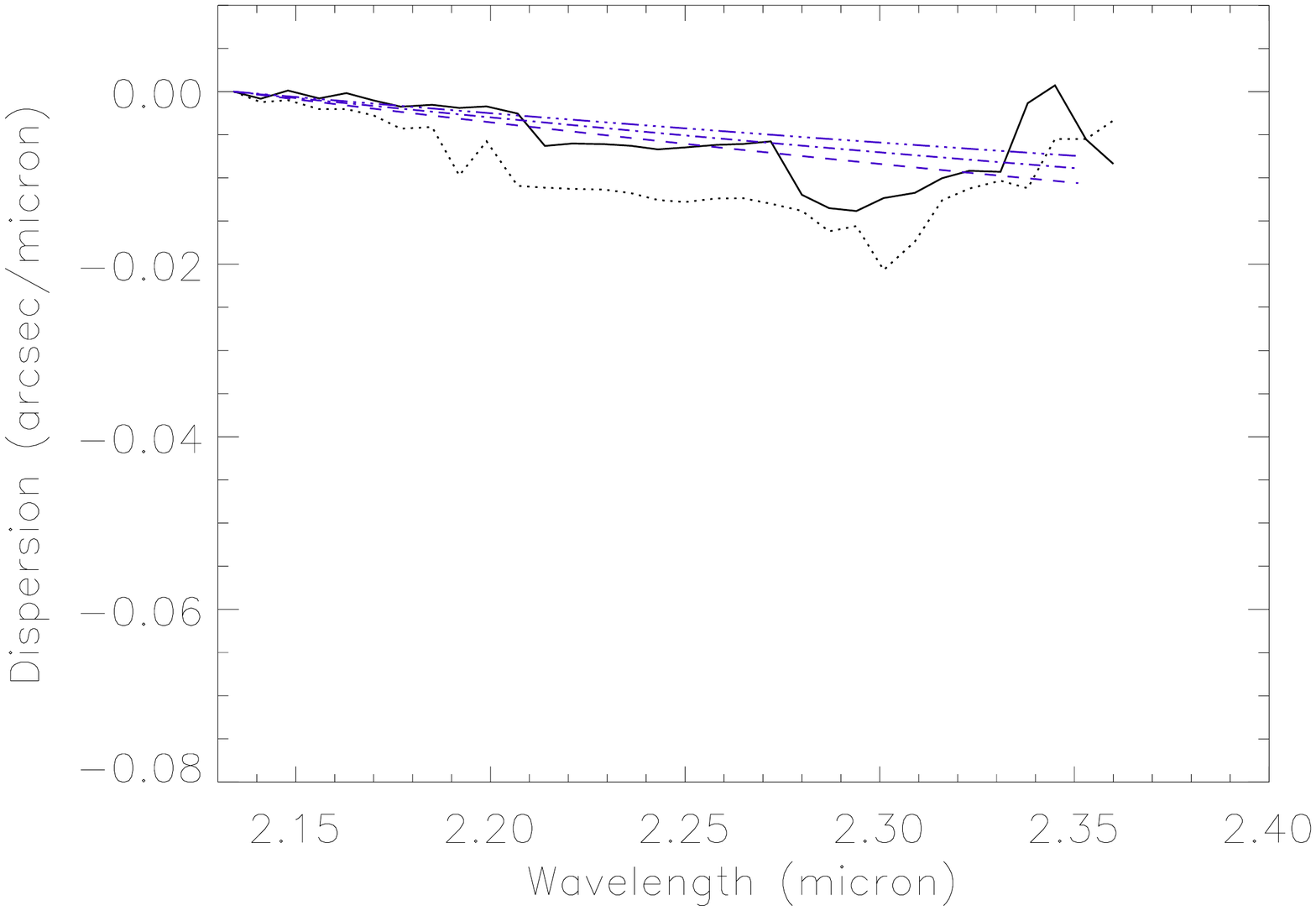} &\includegraphics[scale=0.25]{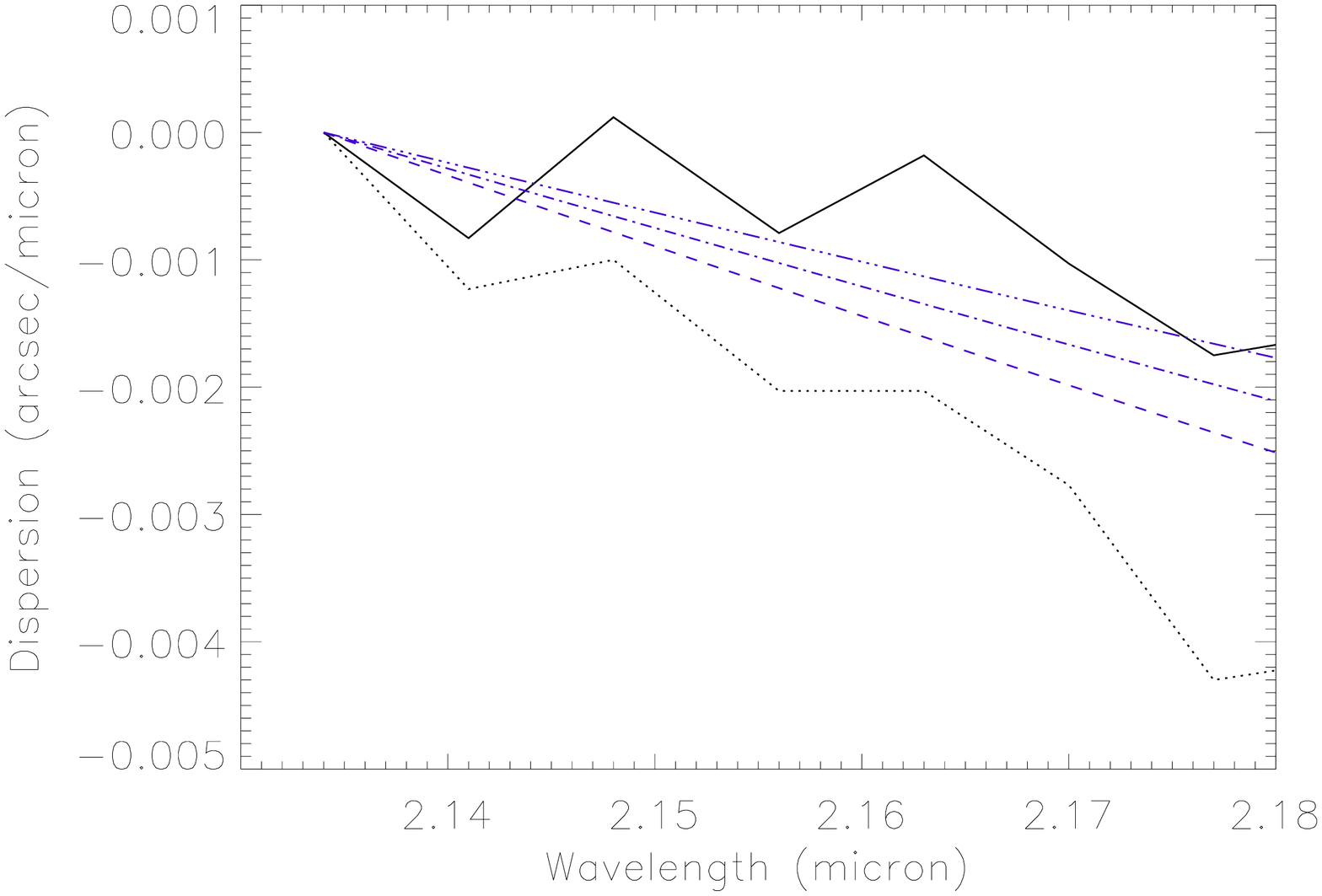}\\
\textit{g.K2-band} & \textit{h. K2-band zoom}\\
\end{tabular}
\caption{Comparison of the ADC dispersion performance at 90mm (dotted line) and 5mm (solid line) separation with the
  simulated atmospheric dispersion at ZD=50deg (dashed line), ZD=45deg (dot-dashed line) and ZD=40deg (dot-dot-dashed line)
  in Y-, J-, H-, and K2-bands. }
\label{dar}
\end{center}
\end{figure}

We compare the dispersion obtained in these data with the one simulated from the script named dar.pro written by Dr. Enrico Marchetti ( ESO, January 2001). This IDL routine computes the Differential Atmospheric Dispersion for a given zenithal distance for different wavelengths with respect to a reference wavelength.
The atmospheric parameters can be adjusted to those characteristic of
the observing site the computation is made for.
Potential errors in the dispersion come from the wavelength calibration in the Y-, J-, H- band filters \cite{schuylerthis}. Due
to the approximation done during the determination of the wavelength
solution, an undefined amount of flux could belong to the next
wavelength instead of the one it has been identified. Moreover, the
centroid determination is accurate to one pixel : 0.0143 arcsec.\\
From Figure~\ref{dar}, we can determine the zenith distance limit for
each filter. 
Laboratory results for the bands K1 and K2 were very noisy and therefore not shown in this paper.
From the laboratory tests, we obtain the following limits :\\
- Y-band : ZD limit = 45 deg\\
- J-band : ZD limit = 40 deg\\
- H-band : ZD limit = 45 deg\\
\pagebreak[4]
\newpage

\subsubsection{Comparison to ZEMAX model}
Our initial investigation of the ADC performance showed that the beam
was displaced on the image by 1.1 arcsec and that the dispersion
direction was perpendicular to its original motion. We then created a
Zemax model that simulated the telescope (no atmosphere, no
repointing) with the ADC. 
We looked at Zemax to check the relative orientation of the prisms including tilts. We were able to verify that when the prisms were rotated relatively by 25 degrees from the “best image” configuration, that the induced astigmatism was consistent with Zemax.
 
We then used laser reflection data to build a new non-sequential Zemax model
that matches the measured behavior. Our first step in
building this model was to try to match the measured data by only
tilting the prisms.  
We had to add a 0.2 deg wedge of the first prism and a 90 deg wedge for the second to match the reflections.
We have checked
that the small angle reflection data also matches this model. In
addition we fed the prism tilts and the wedge from the non-sequential model back into the sequential model to start comparisons of wavefront error and pupil shift.
 
Our latest measurements were planned to be wavefront error, pupil position and IFS images for each of the small-angle rotations we did during the reflection tests. We were able to get data for the -90/-90 case and the -92/-92 case. It appears that the pupil is shifting when the ADC is deployed and this required the input fold mirror plus woofer tip-tilt to point and center the beam on the WFS. 
 

We 
use our current Zemax model to confirm compliance with the
specifications. Values from Macintosh et al.\cite{macintosh2014first} are for a
zenith distance of 30 degrees :\\
- Chromatic pupil shear $<$ 0.4\% :  In Zemax this is calculated as the chromatic blur at the MEMS pupil from 800-2100 nm.\\
- Chromatic aberration $<$ 5 nm across a single band : it includes astigmatism and other low-order
errors. If the Low Order WaveFront Sensor (LOWFS) is operating, these are only relevant between the
science band and H band. We assume an open-loop model can attenuate
the errors by 4x and therefore set the value to 20 nm rms. If the
LOWFS is not operating, the aberrations are relevant only between the
WFS and H band, but we still set the value to 20 nm rms. In Zemax, we
compare the rms WaveFront Error (WFE) at the Focal Plane Mask (FPM) for the center of each wavelength
band. The differences in wavefront maps are subtracted point-by-point
in Matlab to create a residual map.\\
- Chromatic focus offset $<$ 18 nm rms between 800 nm and any science band : This term is driven by the amount of dynamic range in the Adaptve Optics WaveFront Sensor (AOWFS) focus leg.  Currently, this allows 18 nm rms error between 0.8 um and any science band. This could be increased to 36 nm rms by moving the CAL on its bipods. In Zemax we compare the focal position for 800 nm vs the centers of the science bands at the FPM position.\\
The Visible/IR pointing offsets at cass with LOWFS should be less than 30 mas and the ones without LOWFS less than 140 mas :
these are derived for a beam shear of 0.5\% on the MEMS, which corresponds to an intensity contribution of 1e-7 (marginally acceptable) and 2 nm mid-frequency error. The case of 1600 vs YJK1 is for when the LOWFS is in use, and should be less than 30 mas. The case of 800 vs 1600 is only for when the LOWFS is not in use and should be less than 140 mas.\\

\section{On-sky Characterization and Scientific Performance}\label{onsky}
Once GPI was installed on the Gemini South Telescope, and first light
occurred, the ADC was one of the optical systems we needed to
assess.
We first took data with ADC deployed and extracted to check that
the deploy position defined in the laboratory is also valid while on the
telescope. 
We then took data to define the
optimal separation necessary for a target at a zenith distance of
40deg. This test is required for each observing band.
We also need to ensure that the ADC
parameters previously defined are working properly :
at different ZDs,
which also means that the equations defined in the
software are computing correctly the separation needed at
different elevations,
for different target brightness,
 and under different observing conditions, which means
 under good/decent/poor
  seeing and under different percentage of cloud coverage, from clear
  to 70\% cloudy.\\
GPI data were therefore taken during the several commissioning runs, in December 2013,
March 2014 and May 2014. With the datasets obtained, we also checked the header
keywords corresponding to the ADC.
In Table~\ref{stars}, the stars used during the different tests
for the scientific performance and the conditions under which they
were observed are listed. The data reduction was performed using the GPI IDL pipeline \cite{perrinthis}. We used the recipe template
called "Simple Datacube Extraction'' customized to include the mean
combination into 3D datacubes .
\begin{table}[h]
\caption{Target list observed for assessing the GPI ADC scientific performance.} 
\label{stars}
\begin{center}       
\begin{tabular}{|l|l|l|l|l|l|l|l|} 
\hline
\rule[-1ex]{0pt}{3.5ex} Name & Spectral Type & I mag & Jmag &
Hmag & Kmag & Obs. Conditions & Obs. Date (UT) \\
\hline\hline
\rule[-1ex]{0pt}{3.5ex} HIP49404 & A2 & 8.05 & 7.78 & 7.83 & 7.72 &
IQ70 CC50& 2014-03-21\\
\hline
\rule[-1ex]{0pt}{3.5ex} HIP73559 & A8IV & 6.0 & 5.69 & 5.60 & 5.52 & IQ70
CC50 & 2014-03-21\\
\hline
\rule[-1ex]{0pt}{3.5ex} HIP63287 & A1III & 6.86  & 6.64 & 6.65 & 6.61 & IQ70
CC50 & 2014-03-22\\
\hline
\rule[-1ex]{0pt}{3.5ex} Theta1 Ori B & B1V & 8.16 & & 6.30 & 6.00 &
IQ70 CC50 & 2014-03-25\\
\hline
\rule[-1ex]{0pt}{3.5ex} HD142384 & K2III  & 6.53 & 5.57 & 5.11 & 4.87 &
IQ70 CC80  & 2014-05-11\\
\hline
\rule[-1ex]{0pt}{3.5ex} HD95086 & A8III & 7.16 & 6.91 & 6.87 & 6.79 & IQ85
CC70& 2014-05-13\\
\hline
\rule[-1ex]{0pt}{3.5ex} HD114174 & G3IV & 6.0 & 5.61 & 5.31 & 5.20 &
IQ85 CC70 & 2014-05-15\\
\hline
\end{tabular}
\end{center}
\end{table} 
\subsection{On-sky Characterization}
We were able to define the operation procedure
when observing with the ADC deployed. To compensate the pupil shift
introduced when deploying the ADC, we defined a Input Fold
position in the TLC. Therefore, after slewing to the target, and during the
\textit{Align\&Calib} procedure\cite{savranskythis}, the Input Fold moves to this defined
position. The telescope is then re-pointed before taking images.\\
We also verified that the deploy position was well defined in the
software and the separation well determined automatically.\\

\subsection{Throughput}
The datasets useful to compare the throughput we obtained with and
without the ADC deployed have been taken during the March run on different
stars : HIP73559 and HIP63287 (see Table~\ref{stars}).
We used the photometry tool available via the GPI graphical data viewer (GPItv) on the reduced data. The following
parameters were chosen identical to all the datasets : centering box=
5pixels, aperture radius = 5pixels, inner sky radius = 10 pixels,
outer sky radius = 20pixels. The sky algortihm selected is a median sky.


\begin{figure}[!hp]
\begin{center}
\begin{tabular}{c c }
\includegraphics[width=9cm]{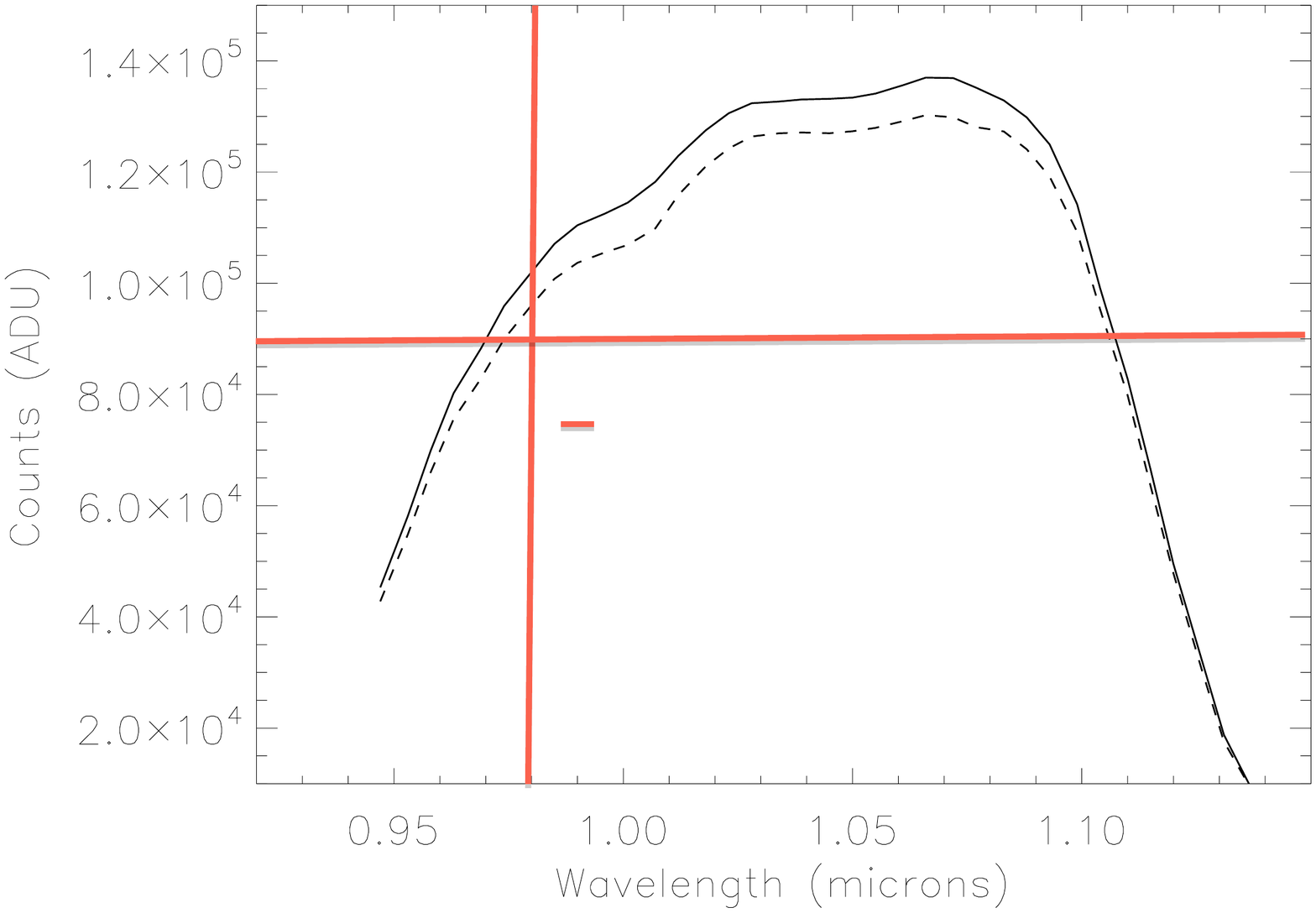} & \includegraphics[width=9cm]{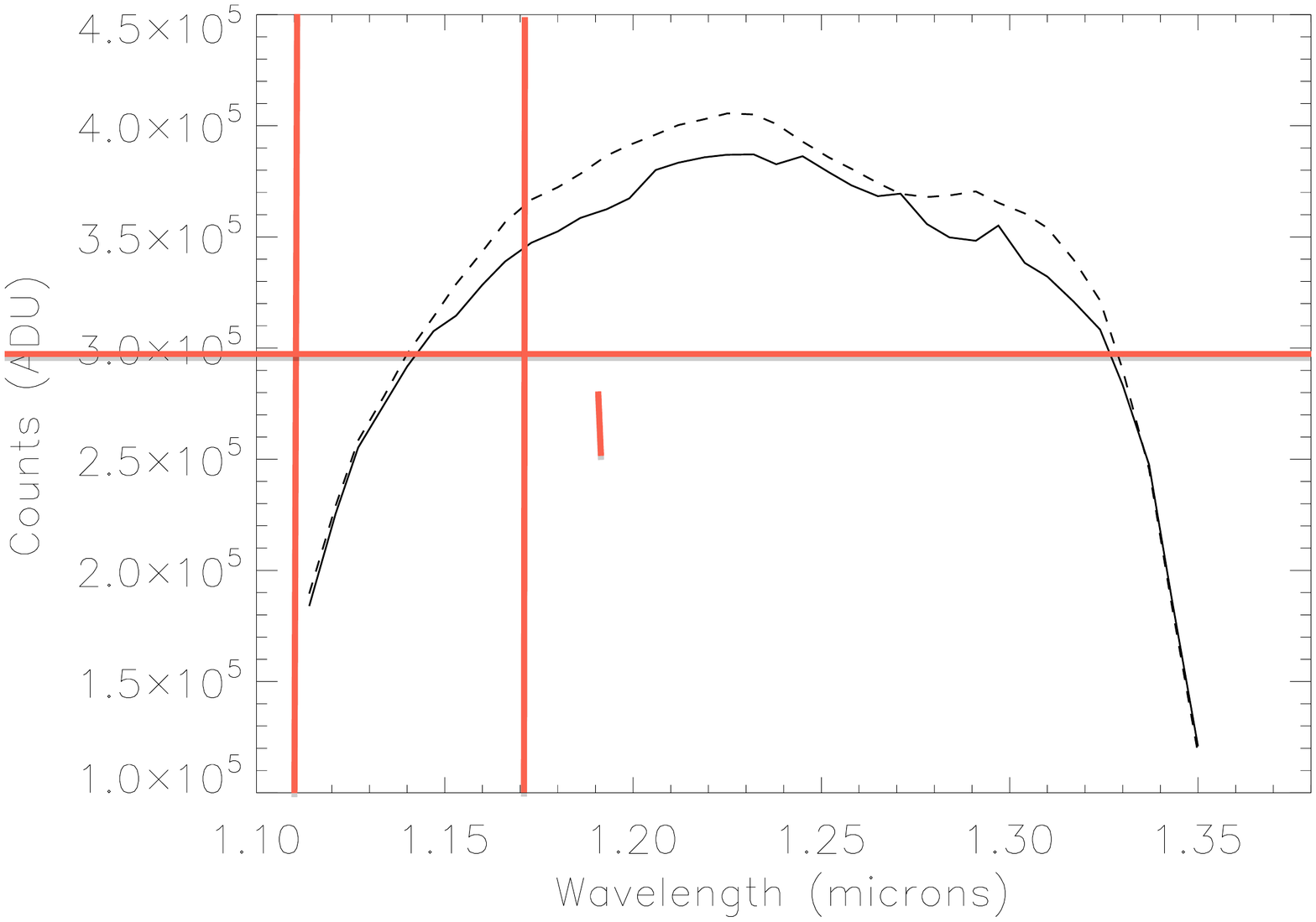}\\
\textrm{a. Y-band} & \textrm{b. J-band}\\
\includegraphics[width=9cm]{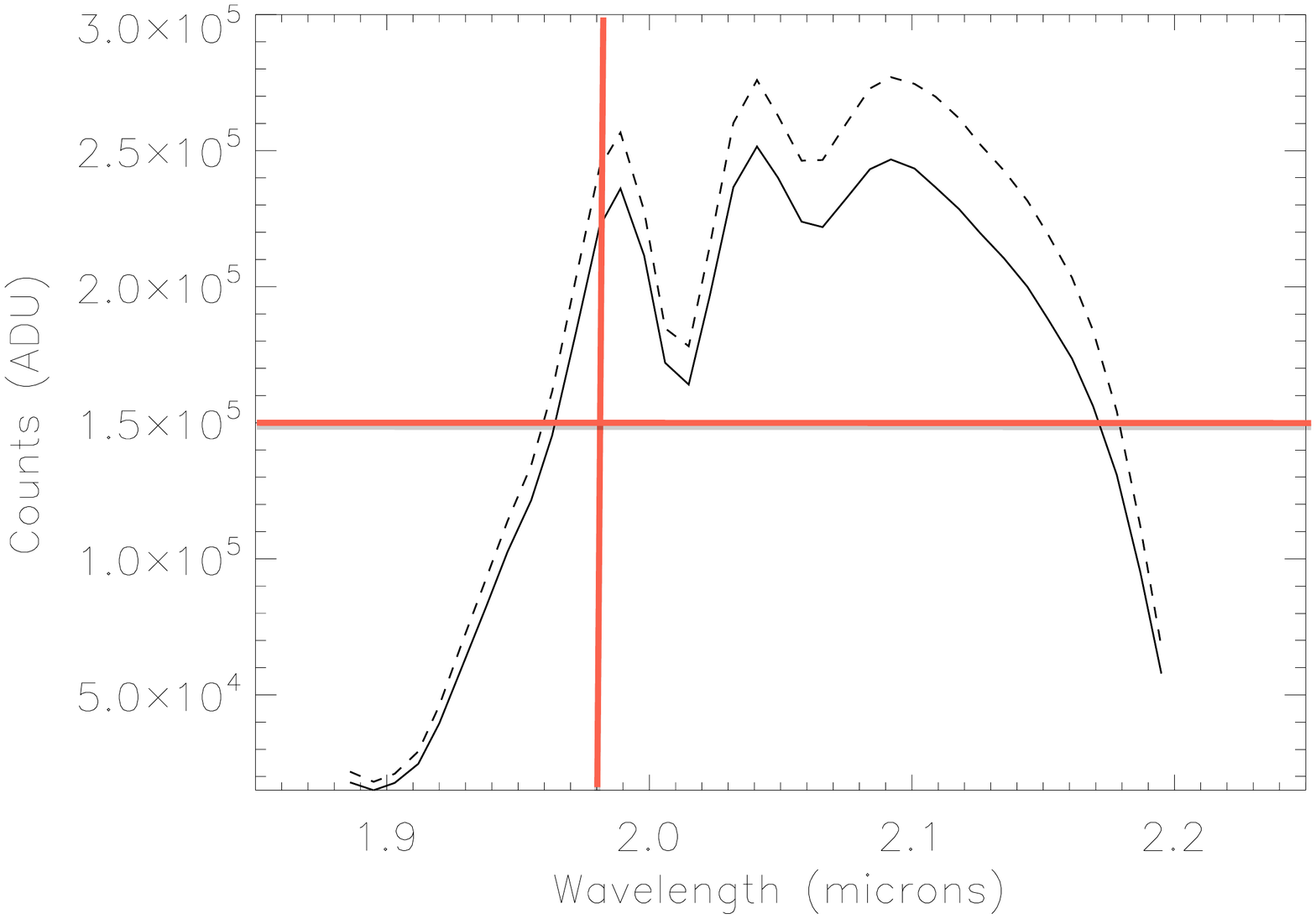} & \\
\textrm{c. K1-band} & \\
\end{tabular}
\caption{Throughput comparison : a. in Y-band, b. in J-band,
  c. K1-band. The solid line represents
    the data with the ADC deployed, and the dashed line the data with
    the ADC extracted.}
\label{throughput}
\end{center}
\end{figure}

The data in J-band (Figure~\ref{throughput}a.) and K1-band
(Figure~\ref{throughput}c.) were taken on the same star, HIP63287, at
ZD=20deg. The data in Y-band (Figure~\ref{throughput}b.) were taken on 
the star, HIP73559, at
ZD=35deg.
In the case of the bluest observing bands, Y- and J-, the throughput
obtained with and without ADC is very similar and we can consider the
influence of the ADC as negligible. 
In the case of the K1-band data, a higher throughput is obtained for
the dataset taken with the ADC extracted. The difference of throughput
reaches 12\%. However, this difference is understandable and expected
as the ADC is aimed to play a major role in the bluer observing bands.

\subsection{Contrast}
\begin{figure}[!hp]
\begin{center}
\begin{tabular}{c c }
\includegraphics[width=9cm]{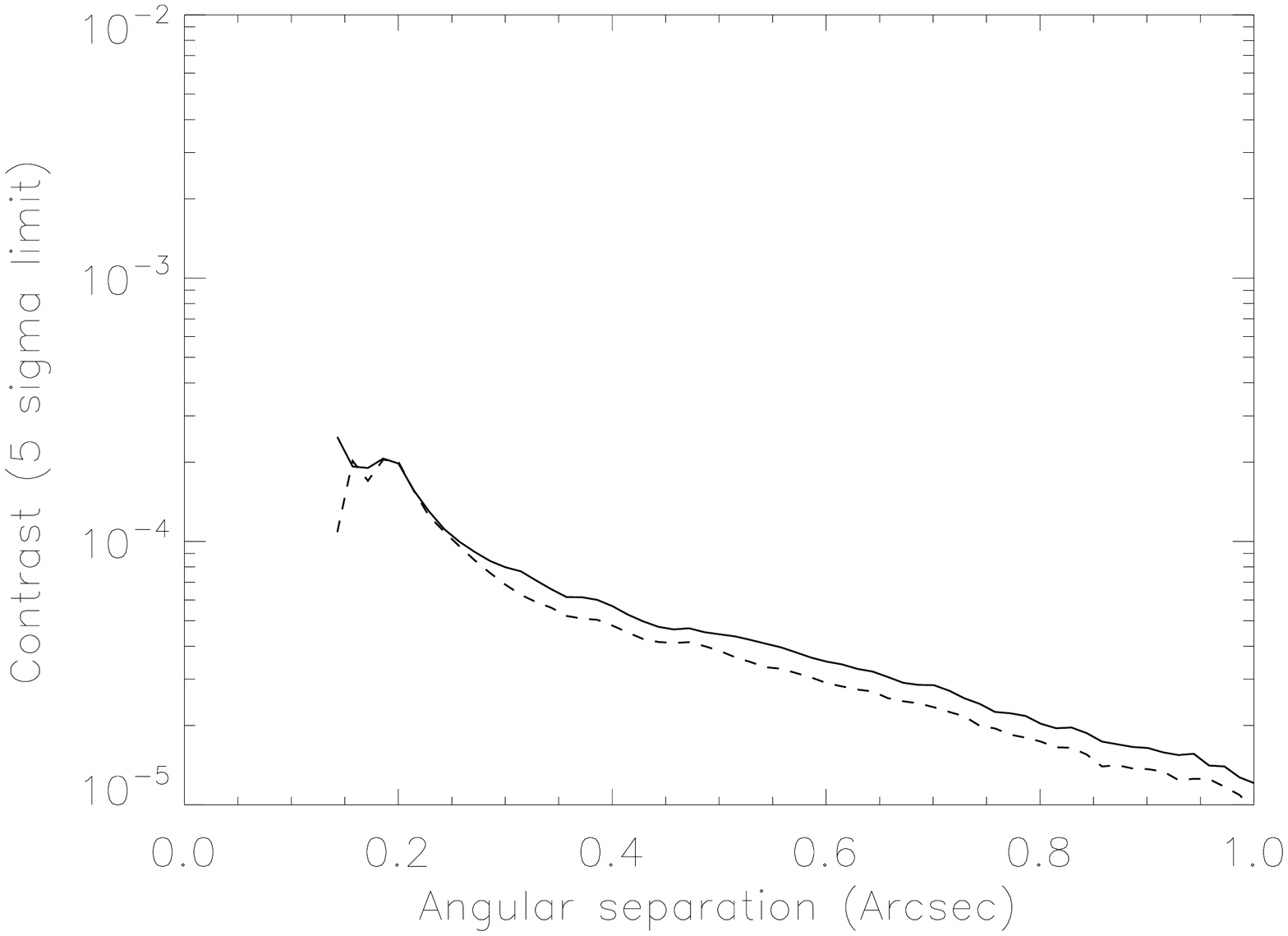} & \includegraphics[width=9cm]{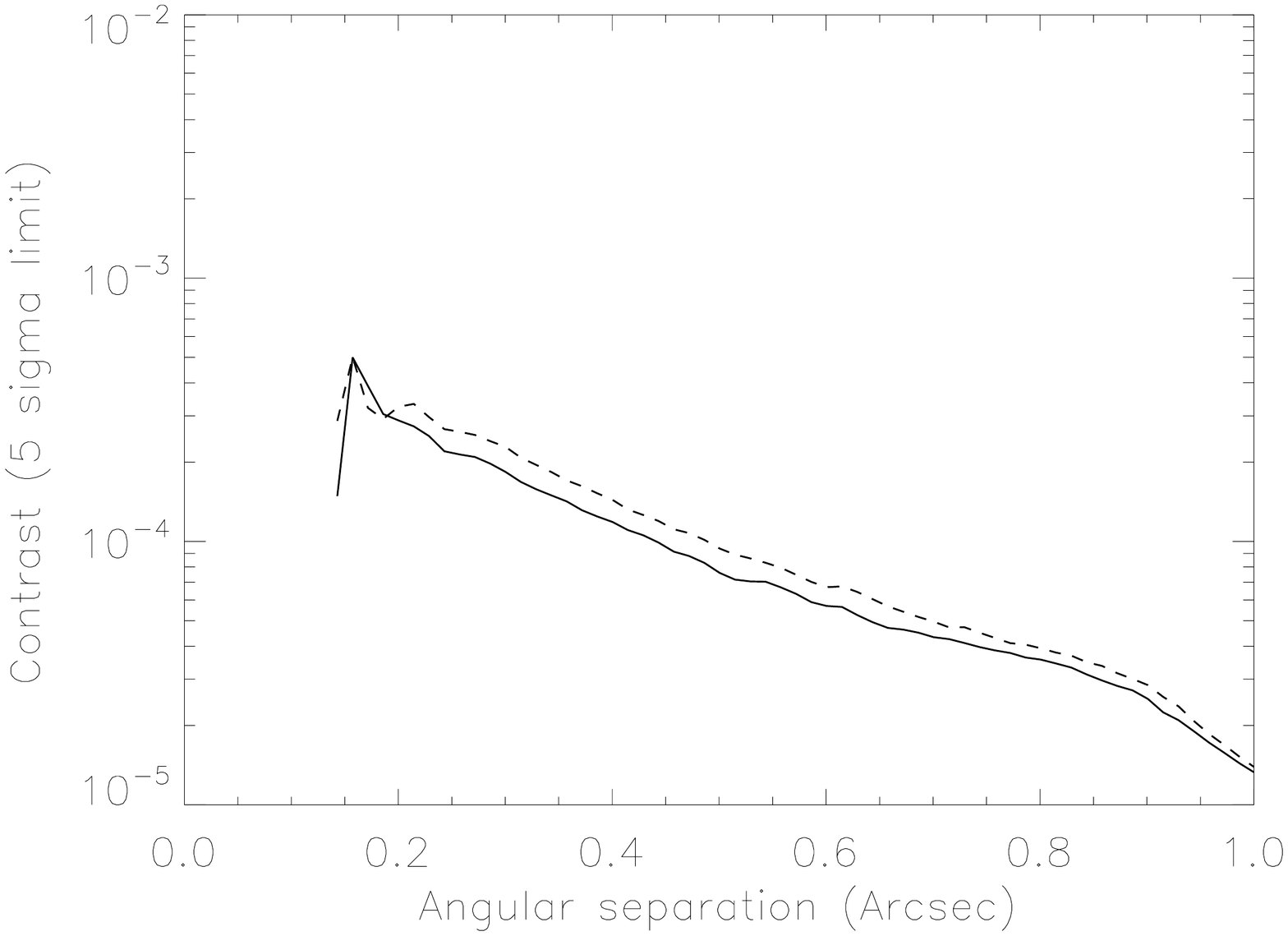}\\
\textrm{a. ZD=20deg} & \textrm{b. ZD=47deg}\\
\end{tabular}
\caption{Comparison of contrast obtained for H-band coronagraphic images with and
    without ADC deployed at ZD=20degrees and ZD=47degrees. The solid line represents
    the data with the ADC deployed, and the dashed line the data with
    the ADC extracted.}
\label{Hconstrast}
\end{center}
\end{figure}

We first collapsed the datacube by Simple Difference Imaging (SDI),
which allows us to significantly attenuate the speckle noise. Details
on this method can be found in Perrin et al\cite{perrinthis}. 
We then removed the
diffuse background light by applying a high-pass filter. 
We obtain the contrast assessment via the GPI graphical data viewer,
GPItv, which computes it
for images of occulted star by using intensities of the satellite
images created by the pupil grid.
\paragraph{Y-band}
We obtained two datasets observed in Y-band coronagraph mode.
The first dataset, on HIP73559, was taken on March 22nd 2014UT. 
The second dataset was taken on May 15th 2014 UT. These observations were
done on a HD114174. Information about the targets and the observing
conditions can be found in Table~\ref{stars}.
Data in Y-band were obtained at zenith distance of 20 degrees and 37 degrees.
Contrast measurements are sensitive to the coronagraph alignment.  
During the Y-band observations at ZD=20deg,
the coronagraph alignment is not optimal. The contrast performance were 
therefore degradated from the misalignment and do not show the true performance 
of the ADC in this wavelength band.
\paragraph{H-band}
We obtained two datasets observed in H-band coronagraph mode.
The first dataset, on HIP73559, was taken on March 22nd 2014UT. 
The second dataset was taken on May 11th 2014 UT. These observations were
done on a HD142384. Information about the targets and the observing
conditions can be found in Table~\ref{stars}.
In Figures~\ref{Hconstrast}, we are comparing
the contrast obtained in two images of 60s exposure time taken at a
zenith distance of 20 degrees and 47 degrees.
Although the dataset obtained in March was taken under clear skies and
good seeing, the May dataset was taken under cloudy skies and good seeing.

\paragraph{K1-band}
We obtained only one dataset using the K1-band coronagraph mode. It
was obtained on HD95086 during the May commissioning run and an
exposure time of 60sec was defined. 
\begin{figure}[!h]
 \centering
  \includegraphics[width=15cm]{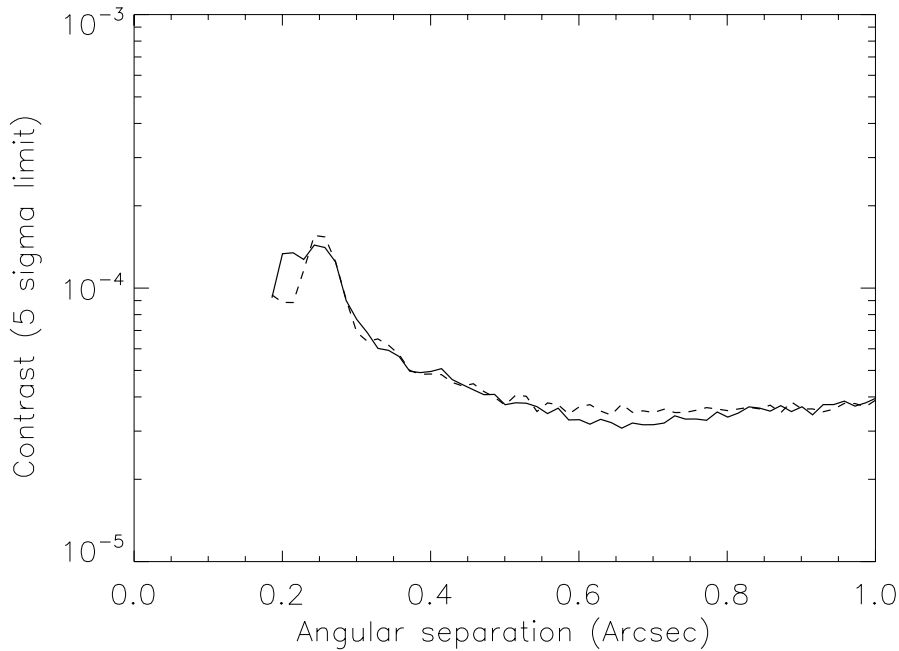}
  \caption{Comparison of contrast obtained for K1-band coronagraphic images with and
    without ADC deployed at ZD=40degrees. The solid line represents
    the data with the ADC deployed, and the dashed line the data with
    the ADC extracted. }
  \label{K1contrastZD40}
\end{figure} 

\subsection{Astrometry}
\begin{figure}[!h]
  \begin{center}
\begin{tabular}{c c}
\includegraphics[width=9cm]{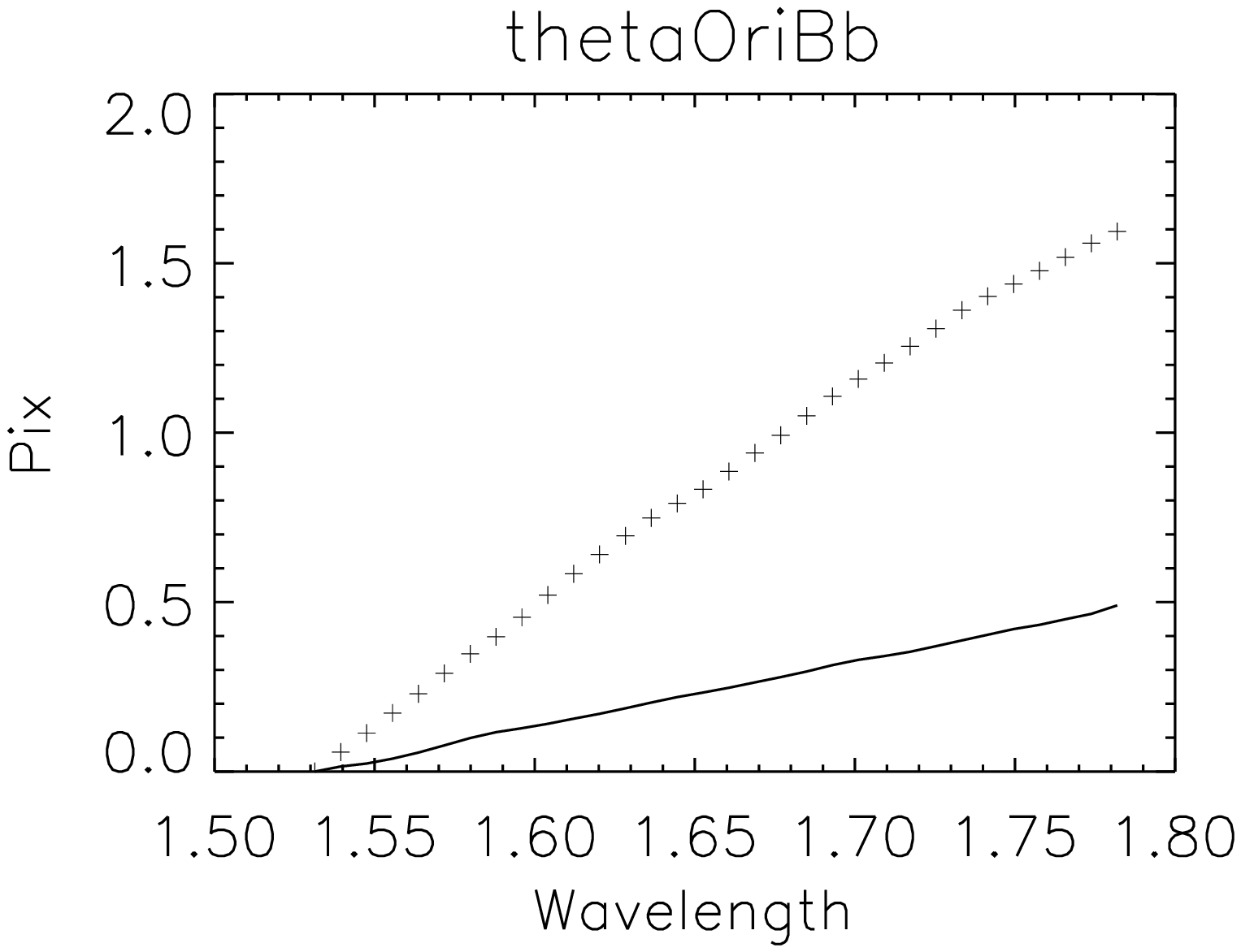} & \includegraphics[width=9cm]{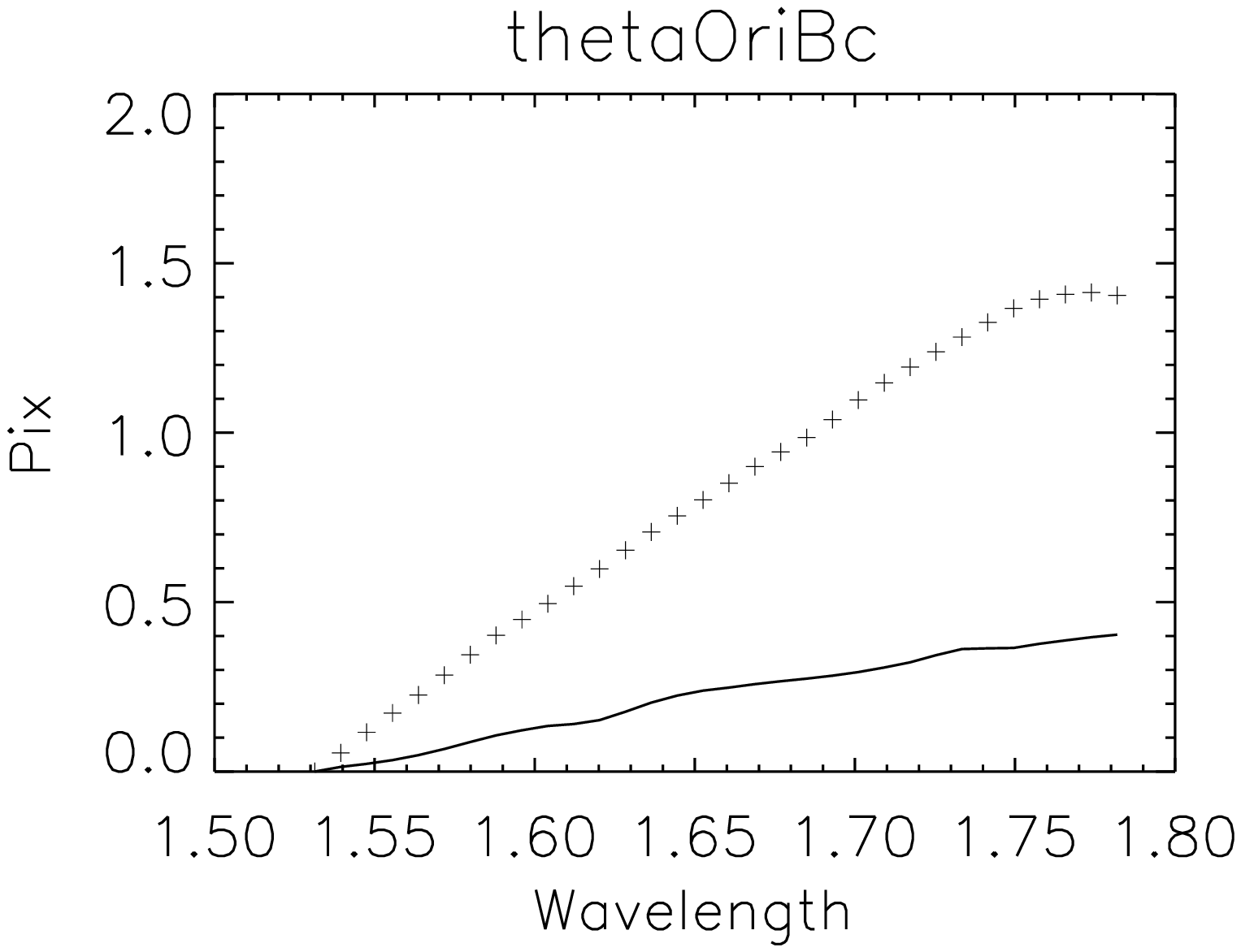}\\ 
\textrm{a. Theta Ori Bb} & \textrm{b. Theta Ori Bc }\\
\includegraphics[width=9cm]{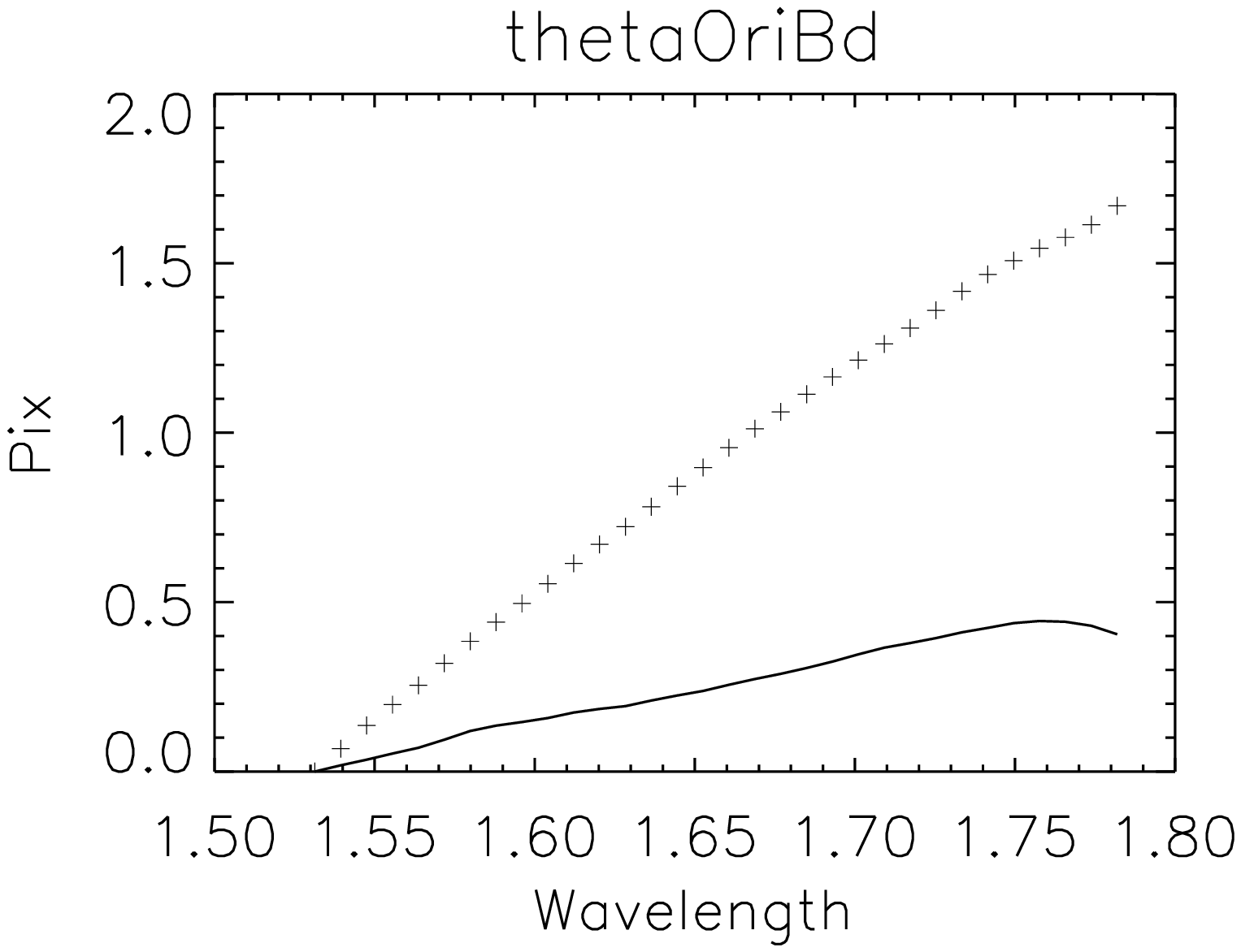} & \\
 \textrm{c. Theta Ori Bd} & \\
\end{tabular}
\caption{Dispersion obtained in the Theta Ori B dataset. The solid line represents the data with the ADC deployed, the dashed line, the dataset with the ADC extracted.}
\label{astrometry2}
\end{center}
\end{figure}

Another way to quantify the validity of the ADC correction is by doing astrometry for a target at larger Zenith distance values.
One of the best targets to measure the astrometry is Theta Ori B. It is bright and has three companions in the GPI field of view. Moreover, during the observing run, we observed the target with a Zenith Distance of 40 degrees for a better dispersion.
The higher the Zenith Distance the more dispersion one can measure without an ADC. This translates in a slope of the pixels position as a function of the wavelength. This is illustrated in Figure~\ref{astrometry2} by the dotted line. \\
In this section, we measure the performance of the ADC by looking at the position of the companions as a function of wavelength. With the ADC deployed, the position of the planet should remain fixed as a function of wavelength. \\
Since the companions are bright relative to the primary star, the satellites spots\cite{wangthis} have a low SNR. A more limiting factor is that one of the satellites falls in between two of the companions. Therefore the satellites spots become the limited factor of the astrometry when the ADC is deployed. We thus looked at the position of the companions relative to the first wavelength.

\begin{table}[h]
\caption{Astrometric results on the photometric field Theta Ori B with and without ADC using the H-band coronagraph observing mode.} 
\label{astrom}
\begin{center}       
\begin{tabular}{|l|l|l|l|} 
\hline
\rule[-1ex]{0pt}{3.5ex} & Theta Ori Bb & Theta Ori Bc & Theta Ori Bd  \\
\hline\hline
\rule[-1ex]{0pt}{3.5ex} Peak to Valley With ADC (mas) & 5.78 & 7.01 & 6.35 \\
\hline
\rule[-1ex]{0pt}{3.5ex} Peak to Valley Without ADC (mas) & 20.2 & 22.79 & 23.88\\
\hline
\rule[-1ex]{0pt}{3.5ex} Standard Deviation With ADC (mas) & 1.78 & 2.04 & 1.97\\
\hline
\rule[-1ex]{0pt}{3.5ex} Standard Deviation Without ADC (mas) & 6.34 & 6.87 & 7.09\\
\hline
\end{tabular}
\end{center}
\end{table} 

As expected, for each companion, the slope is getting closer to zero. However, it is not exactly null. The requirement of the position of the star behind the coronagraph is 5mas and for astrometry 1mas. Table~\ref{astrom} shows that the peak to valley with ADC is within the requirement for the position of the central star. However, the results do not meet the requirements for the astrometry. We could use models to take compensate for the residual errors.
We attribute the remaining slope remaining ADC alignment errors in rotation.
In our errors we also assumed that the AO system was behaving equally for both with and without ADC. This is a valid assumption since the observations with the ADC extracted or deployed have been executed back to back and therefore the seeing has not varied by more than 0.25 arcsec during the observations in and out.

\section{Conclusion}
We characterized the GPI ADC in the laboratory and on the Gemini South Telescope. 
From the on-sky datasets, we realize that the ADC is behaving better
than we expected from the laboratory tests. The zenith distance limits are bigger
than expected.\\
When comparing scientific contrast performance with and without ADC,
we remark that there is no modification of the angular separation
between both configurations. The adaptive optics is therefore
correcting well the effects of the ADC. The astrometric test shows that we do not meet the requirements for the astrometry, which is due to remaining ADC alignment errors in rotation. More data are therefore necessary to continue improving the GPI ADC performance.


\acknowledgments 
The Gemini Observatory is operated by the Association of Universities for Research in
Astronomy, Inc., under a cooperative agreement with the NSF on behalf of the Gemini
partnership: the National Science Foundation (United States), the National Research
Council (Canada), CONICYT (Chile), the Australian Research Council (Australia),
Minist\'erio da Ci\'encia, Tecnologia e Inova\c{c}\=ao (Brazil), and Ministerio de Ciencia,
Tecnolog\'ia e Innovaci\'on Productiva (Argentina).


\bibliography{spie2014}   
\bibliographystyle{spiejour}   




\listoffigures
\listoftables

\end{document}